\documentclass[superscriptaddress,twocolumn,amsmath,amssymb,prd,floatfix,nofootinbib]{revtex4}

\usepackage{graphicx}
\usepackage{dcolumn}
\usepackage[varg]{txfonts}
\usepackage{bm}

\usepackage{color}

\begin{document}

\title{The large-separation expansion of peak clustering in Gaussian
  random fields}

\author{Takahiko Matsubara} \email{tmats@post.kek.jp}
\affiliation{%
  Institute of Particle and Nuclear Studies, High Energy Accelerator
  Research Organization (KEK), Oho 1-1, Tsukuba, Ibaraki 305-0801, Japan
}%
\affiliation{%
  The Graduate University for Advanced Studies (SOKENDAI), Tsukuba,
  Ibaraki 305-0801, Japan
}%
\author{{Sandrine Codis}} \email{codis@iap.fr}
\affiliation{%
CNRS \& Sorbonne Universit\'e, UMR 7095, Institut d'Astrophysique de
Paris, 75014, Paris, France 
  }%
\date{\today}

\begin{abstract}
  In the peaks approach, the formation sites of observable structures
  in the Universe are identified as peaks in the matter density field.
  The statistical properties of the clustering of peaks are
  particularly important in this respect. In this paper, we
  investigate the large-separation expansion of the correlation
  function of peaks in Gaussian random fields. The analytic formula up
  to third order is derived, and the resultant expression can be
  evaluated by a combination of one-dimensional fast Fourier
  transforms, which are evaluated very fast. The analytic formula
  obtained perturbatively in the large-separation limit is compared
  with a method of Monte-Carlo integrations, and a complementarity
  between the two methods is demonstrated.
\end{abstract}

\maketitle


\newcommand{\nuc}{\nu_\mathrm{c}}

\section{\label{sec:Intro}
Introduction
}

All the cosmological structures in the Universe emerge from initial
density fluctuations in the early Universe \cite{bkp}. The formation
of astronomical objects is a complicated process, including nonlinear
dynamics, baryonic physics, radiative transfer, etc., and the relation
between the formation sites of astronomical objects and the spatial
distribution of matter is also complicated in general, the
astronomical objects being biased tracers of the total matter
distribution in the Universe (for a review, see Ref.~\cite{DJS18}).
Modelling this bias relation is one of the key problems in the context
of large-scale structure of the Universe, especially with the advent
of high precision cosmological surveys such as Euclid, LSST, WFIRST to
name a few, which require a detailed modelling of galaxy bias,
flexible and accurate enough not to bias the resulting cosmological
constraints \cite{krause16} -- notably on dark energy, the ultimate
goal of these experiments.

In the peak approach, the formation of dark matter halos is assumed
to take place at the density peaks of the initial density field in
Lagrangian space \cite{Dor70,Kai84,PH85,BBKS}. Although this
assumption is oversimplified, there are empirical evidences that
massive halos corresponds to the high-density peaks in Lagrangian
space \cite{FWDE88,WMS94,Lud11}.

The peak approach can explain the clustering amplitude of rich
clusters, which is much higher than that of galaxies. Indeed, it was
shown that the high-density rare peaks in the Gaussian random field
are more strongly clustered than the underlying (mass) density field
\cite{Kai84}, an effect known as peak bias. Since then, the clustering
properties of peaks in Gaussian random fields have attracted much
attention \cite{BBKS,PW84,Ott86,Cli87,Cat88,LHP89,Col89,RS95,Mat95}.
The biasing by peaks has interesting properties which are not present
in a simple model of local bias. The higher-derivatives of the
underlying density field in the peak formalism affect the scale of
baryon acoustic oscillations \cite{Des08,DCSS10}, and predict
interplay between bias and gravitational evolution \cite{Bal15}. A
combination of the peak approach and the excursion-set approach was
also proposed \cite{AP90,PS12,PSD13,BCDP14} to improve the modelling
of the mass function of dark matter halos.

The evaluation of the correlation function of peaks in the Gaussian
random fields is one of the essential problems in the peak approach.
One can write down an analytic expression for the peak correlation
function in the Gaussian random field, which is represented by
14-dimensional integrals \cite{RS95}. It is possible to adopt
Monte-Carlo integrations to numerically evaluate such integrals at the
price of a large computational cost, as was done in \cite{BCDP16} for
peaks in 1D and in \cite{CPP18} in 3D to predict the connectivity of
the cosmic web. Instead, the correlation function of peaks on large
scales can be evaluated by applying an orthogonal expansion of peaks
\cite{Sza88,Col93,Mat95,LMD16,Diz16}. The orthogonal expansion method
of the Lagrangian bias in Refs.~\cite{Mat95,LMD16} is closely related
to the formalism of integrated Perturbation Theory (iPT)
\cite{Mat11,Mat12,Mat14,MD16}. In fact, the functional coefficients of
the orthogonal expansion are the same as the renormalized bias
functions in terms of iPT for Gaussian initial conditions. These
expansion methods correspond to the large-separation expansion,
because the expansion series are accurate when the correlation
function of the underlying density field is small enough.

In this paper, we explicitly derive analytic formulas for the
large-separation expansion up to third order. While primary
expressions are given with multi-dimensional integrals in Fourier
space, they reduce to combination of one-dimensional Fast Fourier
Transforms (FFTs) in configuration space, applying a technique
developed by Refs.~\cite{SVM16,SV16,MFHB16,FBMH17}. The results are
compared with a Monte-Carlo integration of the full expression of the
peak correlation function.

The paper is organized as follows. We briefly review the peak approach
in Sec.~\ref{sec:Basics}. In Sec.~\ref{sec:Clustering}, a detailed
method to derive the formula for the large-separation expansion is
described, and the formula up to third order is explicitly given. In
the course of the derivation, useful equations to evaluate the bias
coefficients of peaks are presented. In Sec.~\ref{sec:SampleCalc}, a
simple calculation with a cosmological density field is given, and the
result is compared with a Monte-Carlo integration method. Finally,
conclusions are given in Sec.~\ref{sec:Concl}.

\section{\label{sec:Basics}
Basics of peak theory
}

\subsection{\label{subsec:Variables}
Field variables and Gaussian statistics
}

In this section, we review some basic results of peak theory. For
a given density contrast $\delta(\bm{x})$, a smoothed density
contrast is given by
\begin{equation}
  \label{eq:2-1}
  \delta_\mathrm{s}(\bm{x}) = \int d^3x' W_R(|\bm{x}-\bm{x}'|)\,
  \delta(\bm{x}'),
\end{equation}
where $W_R(x)$ is a smoothing kernel, and $R$ is a smoothing radius.
The Fourier transform of the above equation is given by
\begin{equation}
  \label{eq:2-2}
  \tilde{\delta}_\mathrm{s}(\bm{k}) = W(kR)\, \tilde{\delta}(\bm{k}),
\end{equation}
where $\tilde{\delta}_\mathrm{s}(\bm{k})$ and $W(kR)$ are
(3-dimensional) Fourier transforms of $\delta_\mathrm{s}(\bm{x})$ and
$W_R(\bm{x})$ respectively. A common choice for the smoothing kernel
is a Gaussian window, $W(kR) = e^{-k^2R^2/2}$. However, we do not
assume any specific form for the kernel function in this paper, as
long as it filters high-frequency modes.

The power spectrum $P(k)$ of the underlying density field is given by
\begin{equation}
  \label{eq:2-3}
  \left\langle\tilde{\delta}(\bm{k})\,\tilde{\delta}(\bm{k}')\right\rangle
  = (2\pi)^3 \delta_\mathrm{D}^3(\bm{k}+\bm{k}')\,P(k),
\end{equation}
and the power spectrum of the smoothed density field is given by
$P_\mathrm{s}(k) = W^2(kR)\,P(k)$. The spectral moments $\sigma_n(R)$
of the smoothed density field are defined by
\begin{equation}
  \label{eq:2-4}
  {\sigma_n}^2 = \int\frac{k^2dk}{2\pi^2} k^{2n} W^2(kR) P(k).
\end{equation}

With the above notations, the normalized field variables,
\begin{equation}
  \label{eq:2-5}
  \nu(\bm{x}) \equiv \frac{\delta_\mathrm{s}(\bm{x})}{\sigma_0}, \quad
  \eta_i(\bm{x}) \equiv \frac{\partial_i\delta_\mathrm{s}(\bm{x})}{\sigma_1}, \quad
  \zeta_{ij}(\bm{x}) \equiv
  \frac{\partial_i\partial_j\delta_\mathrm{s}(\bm{x})}{\sigma_2},
\end{equation}
are commonly introduced to characterize density peaks. In this paper,
we assume Gaussian statistics for the underlying density field
$\delta$. Because the field variables defined by Eq.~(\ref{eq:2-5})
linearly depend on $\delta$, these variables also obey Gaussian
statistics, i.e., their joint distribution is a multivariate Gaussian.
For a set of variables at a single position, the covariances among the
field variables are given by \cite{BBKS}
\begin{align}
  \label{eq:2-8}
& \langle\nu^2\rangle = 1,\quad
  \langle\nu\eta_i\rangle = 0,\quad
  \langle\nu\zeta_{ij}\rangle =
  -\frac{\gamma}{3}\delta_{ij},\quad
  \langle\eta_i\eta_j\rangle = \frac{1}{3}\delta_{ij},
\nonumber \\                   
&  \langle\eta_i\zeta_{jk}\rangle = 0,\quad
  \langle\zeta_{ij}\zeta_{kl}\rangle =
  \frac{1}{15}
  (\delta_{ij}\delta_{kl}+\delta_{ik}\delta_{jl}+\delta_{il}\delta_{jk}),
\end{align}
where
\begin{equation}
  \label{eq:2-9}
  \gamma \equiv \frac{{\sigma_1}^2}{\sigma_0\sigma_2}
\end{equation}
characterizes the broadband shape of the smoothed power spectrum of
the underlying density field.

Because the matrix of (rescaled) second derivatives, $\zeta_{ij}$, is
a symmetric tensor, only six components with $i\geq j$ are
independent. Therefore, we have ten independent variables
defined in Eq.~(\ref{eq:2-5}) at each position $\bm{x}$. We
denote this 10-dimensional set of variables as $\bm{y}$ at
each position, i.e.,
\begin{equation}
  \label{eq:2-10}
  \bm{y} \equiv \left(\nu,\eta_1,\eta_2,\eta_3,
    \zeta_{11},\zeta_{22},\zeta_{33},\zeta_{12},\zeta_{23},\zeta_{13}\right).
\end{equation}
The probability distribution function at a single point is thus given
by
\begin{align}
  \label{eq:2-11}
  \mathcal{P}(\bm{y})
  &=
  \frac{1}{\sqrt{(2\pi)^{10}\det M}}
  \exp\left(-\frac{1}{2}\bm{y}^\mathrm{T} M^{-1} \bm{y}\right)
    \nonumber\\
  &\propto
  \exp\left[-\frac{\nu^2 + {J_1}^2 - 2\gamma\nu J_1}{2(1-\gamma^2)}
    - \frac{3}{2} \eta^2 - \frac{5}{2} J_2 \right],
\end{align}
where $M$ is the covariance matrix of $\bm{y}$, whose components are
given by $M_{\alpha\beta} = \langle y_\alpha y_\beta \rangle$. In the
last expression, we have also used rotationally invariant quantities
\cite{Dor70,PGP09,GPP12}
\begin{equation}
  \label{eq:2-12}
  \eta^2 \equiv \bm{\eta}\cdot\bm{\eta}, \quad
  J_1 \equiv -\zeta_{ii}, \quad
  J_2 \equiv \frac{3}{2} \tilde{\zeta}_{ij} \tilde{\zeta}_{ji}, \quad
  J_3 \equiv \frac{9}{2} \tilde{\zeta}_{ij} \tilde{\zeta}_{jk} \tilde{\zeta}_{ki},
\end{equation}
where repeated indices are summed over, and
\begin{equation}
  \label{eq:2-13}
  \tilde{\zeta}_{ij} \equiv \zeta_{ij} + \frac{1}{3} \delta_{ij} J_1,
\end{equation}
is the traceless part of $\zeta_{ij}$. It is a consequence of the
rotational symmetry of the statistics that the distribution function
should only depend on rotationally invariant quantities. One
should note that the probability distribution function of
Eq.~(\ref{eq:2-11}) is for the linear variable $\bm{y}$ even in the
last expression, and that $\eta^2$, $J_2$, $J_3$ are nonlinear
functions of the field derivatives $\eta_i$ and $\zeta_{ij}$.

\subsection{\label{subsec:NumberDensity}
The number density of peaks
}

The number density of peaks with height between $\nuc$ and
$\nuc+d\nuc$ is denoted by $n_\mathrm{pk}d\nuc$, where
$n_\mathrm{pk}(\nuc,\bm{x})$ is the differential number density of
peaks at a given position $\bm{x}$. This quantity is itself a random
variable and can be expressed in terms of the density field and its
derivatives. The expression is derived by Taylor expanding the density
field close to a local extremum where $\eta_i= 0$ \cite{Kac43,Ric44}.
As a result, the differential number density $n_\mathrm{pk}$ at
position $\bm{x}$ is given by \cite{BBKS}
\begin{equation}
  \label{eq:2-14}
  n_\mathrm{pk}(\nuc,\bm{x}) = 
  \frac{3^{3/2}}{{R_*}^3} \delta_\mathrm{D}(\nu-\nuc) 
  \delta_\mathrm{D}^3(\bm{\eta})\Theta(\lambda_3)|\det\bm{\zeta}|,
\end{equation}
where $R_*\equiv \sqrt{3}\sigma_1/\sigma_2$ is the characteristic
radius of a peak, and $\lambda_3$ is the smallest eigenvalue of the
$3\times 3$ matrix $(-\zeta_{ij})$. Eq.~(\ref{eq:2-14}) imposes the
height of the peak, a zero gradient and all eigenvalues to be positive
with a Jacobian $|\det\bm{\zeta}|$ that corresponds to the typical
volume associated with a peak.

Relying on ergodicity, the (spatial) average of the differential
number density is calculated by taking the ensemble average of
Eq.~(\ref{eq:2-14}) with the probability distribution function of
Eq.~(\ref{eq:2-11}), which yields \cite{BBKS}
\begin{equation}
  \label{eq:2-15}
  \bar{n}_\mathrm{pk}(\nuc) =
  \frac{1}{(2\pi)^{3/2} {R_*}^3}
  \int_0^\infty dx\,f(x)\,\mathcal{N}(\nu,x),
\end{equation}
where
\begin{equation}
  \label{eq:2-15-1}
  \mathcal{N}(\nu,J_1) =
  \frac{1}{2\pi\sqrt{1 - \gamma^2}}
  \exp\left[-\frac{\nu^2 + {J_1}^2 - 2\gamma\nu J_1}{2(1-\gamma^2)}\right]
\end{equation}
is the joint distribution function of $\nu$ and $J_1$, and
\begin{multline}
  \label{eq:2-16}
  f(x) =
  \frac{x}{2}\left(x^2-3\right)
  \left[
    \mathrm{erf}\left(\frac{1}{2}\sqrt{\frac{5}{2}}\,x\right) 
    + \mathrm{erf}\left(\sqrt{\frac{5}{2}}\,x\right) 
  \right]
  \\
  + \sqrt{\frac{2}{5\pi}}
  \left[
    \left(\frac{x^2}{2} - \frac{8}{5}\right) e^{-5x^2/2}
    + \left(\frac{31}{4}x^2 + \frac{8}{5}\right) e^{-5x^2/8}
  \right].
\end{multline}

The number density $N_\mathrm{pk}$ of peaks above a height $\nuc$ can
be derived by replacing the delta function
$\delta_\mathrm{D}(\nu-\nuc)$ by the step function $\Theta(\nu-\nuc)$
in Eq.~(\ref{eq:2-14}). Equivalently, we have
\begin{equation}
  \label{eq:2-17}
  N_\mathrm{pk}(\nuc,\bm{x}) =
  \int_{\nuc}^\infty d\nuc'\,
  n_\mathrm{pk}(\nuc',\bm{x}).
\end{equation}
Taking the average of the above equation, and using
Eq.~(\ref{eq:2-15}), we get
\begin{equation}
  \label{eq:2-18}
  \bar{N}_\mathrm{pk}(\nuc) = \frac{1}{2(2\pi)^2 {R_*}^3}
  \int_0^\infty dx\,f(x)\,e^{-x^2/2}\,
  \mathrm{erfc}
  \left[
    \frac{\nuc-\gamma x}{\sqrt{2(1-\gamma^2)}}
  \right].
\end{equation}
However, a physical selection criterion for peaks which would form a
given class of objects is unlikely to be so sharp \cite{BBKS}. Instead
of the step function, one may generally consider an increasing function
$\Xi(\nu,\nuc)$ which satisfies
$\Xi(-\infty,\nuc) = 0$, $\Xi(+\infty,\nuc) = 1$,
and the transition from zero to one occurs around the threshold value
$\nuc$. The function $\Xi(\nu,\nu_c)$ corresponds to the
function $t(\nu/\nuc)$ in Ref.~\cite{BBKS}. In this case,
Eqs.~(\ref{eq:2-17}) and (\ref{eq:2-18}) are replaced by
\begin{align}
  \label{eq:2-19}
  N_\mathrm{pk}(\nuc,\bm{x})
  &=
  \int_{-\infty}^\infty d\nuc'\, \Xi(\nuc',\nuc)\,
    n_\mathrm{pk}(\nuc',\bm{x})
    \nonumber\\
  &= 
  \frac{3^{3/2}}{{R_*}^3} \Xi(\nu,\nuc) 
  {\delta_\mathrm{D}}^3(\bm{\eta})\Theta(\lambda_3)|\det\bm{\zeta}|,
\end{align}
and
\begin{multline}
  \label{eq:2-20}
  \bar{N}_\mathrm{pk}(\nuc) =
  \frac{1}{(2\pi)^2 {R_*}^3\sqrt{2\pi(1-\gamma^2)}}
  \int_0^\infty dx\,f(x)\,e^{-x^2/2}
  \\
  \times
  \int_{-\infty}^\infty d\nuc'\, \Xi(\nuc',\nuc)\,
  \exp\left[-\frac{(\nuc'-\gamma x)^2}{2(1-\gamma^2)}\right].
\end{multline}

\section{\label{sec:Clustering}
Clustering of peaks
}

\subsection{\label{subsec:Expansion}
Clustering of peaks and the renormalized bias functions
}

The clustering of peaks can be characterized by their correlation
functions. In particular, the two-point correlation function
$\xi_\mathrm{pk}(r)$ of density peaks above a threshold $\nuc$ is
defined by
\begin{equation}
  \label{eq:3-21}
  1 + \xi_\mathrm{pk}\left(|\bm{x}_1-\bm{x}_2|\right) =
  \frac{
  \left\langle
    N_\mathrm{pk}(\nuc,\bm{x}_1)
    N_\mathrm{pk}(\nuc,\bm{x}_2)
  \right\rangle}
  {{\bar{N}_{\mathrm{pk}}}^2(\nuc)}.
\end{equation}
Similarly, one can consider $N$-point correlation functions in
general. For simplicity, we calculate only the two-point correlation
function in this paper, while higher-order functions can be evaluated
using a similar technique as developed below.

The three-dimensional Fourier transform of the correlation function is
the power spectrum, $P_\mathrm{pk}(k)$. For statistically homogeneous
and isotropic field, they are related by
\begin{align}
  \label{eq:3-22a}
  P_\mathrm{pk}(k) &= 4\pi \int r^2 dr j_0(kr) \xi_\mathrm{pk}(r),
  \\
  \label{eq:3-22b}
  \xi_\mathrm{pk}(r) &= \int \frac{k^2dk}{2\pi^2} j_0(kr) P_\mathrm{pk}(k).
\end{align}

The clustering of peaks are fully determined by the statistics of the
underlying density field and its derivatives up to second order. The
statistics of a Gaussian density field is uniquely determined by its
power spectrum. In our case, the power spectrum of the underlying
density field is given by $P_\mathrm{s}(k) = W^2(kR) P(k)$. Thus, the
power spectrum of peaks, $P_\mathrm{pk}(k)$, is considered as a
functional of $P_\mathrm{s}(k)$. The number density of peaks is a
specific example of a biased field. There is a systematic way of
expanding the biased power spectrum in terms of the underlying power
spectrum \cite{Mat95,Mat11,MD16}. In the absence of dynamical
evolution, the power spectrum of peaks has a form,
\begin{equation}
  \label{eq:3-23}
  P_\mathrm{pk}(k) =
  \sum_{n=1}^\infty \frac{1}{n!}
  \int_{\bm{k}_{1\cdots n}=\bm{k}}
  \left[
    c_n\left(\bm{k}_1,\ldots,\bm{k}_n\right)
  \right]^2
  P_\mathrm{s}(k_1)\cdots P_\mathrm{s}(k_n),
\end{equation}
where we adopt the notation
\begin{equation}
  \label{eq:3-23-1}
  \int_{\bm{k}_{1\cdots n}=\bm{k}} \cdots \equiv
  \int \frac{d^3k_1}{(2\pi)^3} \cdots \frac{d^3k_n}{(2\pi)^3}
  (2\pi)^3\delta_\mathrm{D}^3(\bm{k}_{1\cdots n}-\bm{k}) \cdots
\end{equation}
and $\bm{k}_{1\cdots n}\equiv \bm{k}_1 + \cdots + \bm{k}_n$. The
appearance of the delta function in the integral is a consequence of
the translational invariance of space.

The functions $c_n(\bm{k}_1,\ldots,\bm{k}_n)$ are called ``the
renormalized bias functions'' in the formalism of iPT \cite{Mat11}, and are defined as
\begin{multline}
  \label{eq:3-24}
  \left\langle
    \frac{\delta^n \tilde{N}_\mathrm{pk}(\nuc,\bm{k})}
    {\delta\delta_\mathrm{s}(\bm{k}_1)\cdots\delta\delta_\mathrm{s}(\bm{k}_n)}
  \right\rangle
  \\=
  \frac{\bar{N}_\mathrm{pk}(\nuc)}{(2\pi)^{3n}}
  (2\pi)^3\delta_\mathrm{D}^3(\bm{k}_{1\cdots n}-\bm{k})
  c_n(\bm{k}_1,\ldots,\bm{k}_n),
\end{multline}
in the case of peaks, where $\tilde{N}_\mathrm{pk}(\nuc,\bm{k})$ is
the Fourier transform of $N_\mathrm{pk}(\nuc,\bm{x})$, and
$\delta/\delta\delta_\mathrm{s}(\bm{k})$ denotes the functional
derivative with respect to $\delta_\mathrm{s}(\bm{k})$. The
renormalized bias functions can also be considered as the coefficients
of a generalized Wiener-Hermite expansion \cite{Mat95}, and are akin
to the multipoint propagators \cite{BCS08} in the context of
cosmological perturbation theory. Note that Eq.~(\ref{eq:3-23}) holds
only for Gaussian density field.

\subsection{\label{subsec:BiasFunctions}
Renormalized bias functions and bias coefficients for peaks
}

The renormalized bias functions of peaks up to third order are first
inferred by an analogy with the so-called local bias approach
\cite{Des13,LMD16}. The same results are shown to be directly derived
from the original definition of the renormalized bias function,
Eq.~(\ref{eq:3-24}), up to second order \cite{MD16}. In
Appendix~\ref{app:cthree}, we show the third-order renormalized bias
functions can also be directly derived from the definition. The
results are given by
\begin{align}
  \label{eq:3-25a}
  &c_1(\bm{k}) = b_{10} + b_{01}k^2
  \\
  \label{eq:3-25b}
  &c_2(\bm{k}_1,\bm{k}_2)
  = b_{20} + b_{11}\left({k_1}^2 + {k_2}^2\right)
    + b_{02} {k_1}^2 {k_2}^2
    \nonumber\\
  & \hspace{4pc} - 2 \chi_1 (\bm{k}_1\cdot\bm{k}_2)
  + \omega_{10}
    \left[ 3 (\bm{k}_1\cdot\bm{k}_2)^2 - {k_1}^2 {k_2}^2\right]
  \\
  \label{eq:3-25c}
  &c_3(\bm{k}_1,\bm{k}_2,\bm{k}_3)
    = b_{30} + b_{21}\left({k_1}^2 + {k_2}^2 + {k_3}^2\right)
    \nonumber\\
  & \hspace{2pc}
    + b_{12}\left( {k_1}^2 {k_2}^2 + \mathrm{cyc.}\right)
  + b_{03} {k_1}^2 {k_2}^2 {k_3}^2
    \nonumber\\
  & \hspace{2pc}
  - 2 b_{10} \chi_1 (\bm{k}_1\cdot\bm{k}_2 + \mathrm{cyc.})
  - 2 b_{01} \chi_1
    \left[(\bm{k}_1\cdot\bm{k}_2){k_3}^2 + \mathrm{cyc.}\right]
    \nonumber\\
  & \hspace{2pc}
  + c_{10010}
    \left\{\left[ 3 (\bm{k}_1\cdot\bm{k}_2)^2 - {k_1}^2 {k_2}^2\right]
    + \mathrm{cyc.}\right\}
    \nonumber\\
  & \hspace{2pc}
  + 3 c_{01010}
    \left[(\bm{k}_1\cdot\bm{k}_2)^2{k_3}^2 + \mathrm{cyc.}
    - {k_1}^2 {k_2}^2 {k_3}^2\right]
    \nonumber\\
  & \hspace{2pc}
  - 3 \varpi_{01}
    \Biggl[
    (\bm{k}_1\cdot\bm{k}_2)(\bm{k}_2\cdot\bm{k}_3)(\bm{k}_3\cdot\bm{k}_1)
    \nonumber\\
  & \hspace{5pc}
  - \frac{1}{3} (\bm{k}_1\cdot\bm{k}_2)^2 {k_3}^2 + \mathrm{cyc.}
    + \frac{2}{9} {k_1}^2 {k_2}^2 {k_3}^2
    \Biggr],
\end{align}
where ``$+\,\mathrm{cyc.}$'' indicates additions of cyclic
permutations of the previous term. The coefficients are defined by
\begin{equation}
  \label{eq:3-26}
  b_{ij} = c_{ij000}, \quad
  \chi_q = c_{00q00}, \quad
  \omega_{lm} = c_{000lm}, \quad
  \varpi_{01} = \frac{45}{\sqrt{7}} \omega_{01},
\end{equation}
and $c_{ijqlm}$ are generic bias coefficients which were introduced by
Ref.~\cite{LMD16} in the context of an orthonormal expansion of the
biased field based on rotationally invariant polynomials. They are
defined by
\begin{multline}
  \label{eq:3-27}
  c_{ijqlm}(\nuc)
  = (-1)^q
  \\ \times
  \frac{\left\langle N_\mathrm{pk}(\nuc)\,H_{ij}(\nu,J_1)\,
      L^{(1/2)}_q\left(3\eta^2/2\right)\,
      F_{lm}\left(5J_2, J_3\right)\right\rangle}
    {{\sigma_0}^i\,{\sigma_1}^{2q}\,{\sigma_2}^{j+2l+3m}\,
      \bar{N}_\mathrm{pk}(\nuc)},
\end{multline}
where $\langle\cdots\rangle = \int d^{10}y \cdots \mathcal{P}(\bm{y})$
is the ensemble average of the field variables. The functions $H_{ij}$
and $L^{(1/2)}_q$ are multivariate Hermite polynomials and generalized
Laguerre polynomials, respectively, and $F_{lm}$ are orthogonal
polynomials. The definitions and orthogonality relations of these
polynomials are given in Appendix~\ref{app:polynomials}.

Since we have $N_\mathrm{pk} \propto \delta_\mathrm{D}^3(\bm{\eta})$,
the generalized Laguerre polynomials $L^{(1/2)}_q(3\eta^2/2)$ in
Eq.~(\ref{eq:3-27}) can be replaced by $L^{(1/2)}_q(0)$ and the factor
$\chi_q$ is always factored out as
\begin{equation}
  \label{eq:3-28}
  c_{ijqlm} = c_{ij0lm} \chi_q.
\end{equation}
However, in general, $c_{ijqlm} \neq b_{ij}\chi_q\omega_{lm}$.
Because $L^{(\alpha)}_q(0) =
\Gamma(q+\alpha+1)/[\Gamma(q+1)\Gamma(\alpha+1)]$, the coefficient
$\chi_q$ is explicitly given by
\begin{equation}
  \label{eq:3-29}
  \chi_q =
  \frac{(-1)^q\,\Gamma(q+3/2)}{q!\,\Gamma(3/2)\,{\sigma_1}^{2q}} =
  \frac{(-1)^q\,(2q+1)!!}{2^q\,q!\,{\sigma_1}^{2q}}.
\end{equation}

Although the expression in the right-hand side of Eq.~(\ref{eq:3-27})
is an explicit ten-dimensional integral, it can be evaluated by simple
combinations of one-dimensional integrals. The derivation is given in
Appendix~\ref{app:coeff}, and the results are
\begin{equation}
  \label{eq:3-30}
  c_{ijqlm}(\nuc) =
  \frac{\chi_q}{{\sigma_0}^i{\sigma_2}^{j+2l+3m}}
  \frac{\int_0^\infty dx\,
    \mathcal{M}_{ij}(\nuc,x)\,f_{lm}(x)}
  {\int_0^\infty dx\,
    \mathcal{M}_{00}(\nuc,x)\,f_{00}(x)},
\end{equation}
where
\begin{align}
  \label{eq:3-31a}
  &\mathcal{M}_{ij}(\nuc,x)
  \equiv
  \int d\nu\,\Xi(\nu,\nuc) \mathcal{N}(\nu,x)\,
    H_{ij}(\nu,x),
\\
  \label{eq:3-31b}
  &f_{lm}(x)
  \equiv
  \frac{3^2 5^{5/2}}{\sqrt{2\pi}}
  \left(
    \int_0^{x/4} dy \int_{-y}^y dz 
    + \int_{x/4}^{x/2} dy \int_{3y-x}^y dz 
  \right)
  \nonumber\\
  & \quad \times e^{-5(3y^2+z^2)/2}
  G(x,y,z)\,F_{lm}\left(15y^2+5z^2,z^3-9y^2z\right),
\end{align}
and
\begin{equation}
  \label{eq:3-32}
  G(x,y,z) \equiv
  (x-2z)\left[(x+z)^2 - (3y)^2\right]y(y^2-z^2).
\end{equation}

When the function $\Xi$ is a simple function, such as a step function,
the integration of $\mathcal{M}_{ij}$ may be analytically performed.
Otherwise, for a fixed value of the threshold $\nuc$, the functions
$\mathcal{M}_{ij}$ can be evaluated by one-dimensional integrations, and
tabulated for further integrations of Eq.~(\ref{eq:3-30}).

The last function $G(x,y,z)$ is denoted as $F(x,y,z)$ in
Ref.~\cite{BBKS}. The integral of $f_{lm}$ can be analytically
performed for each integers $l$ and $m$. The function $f_{00}(x)$ is
the same as $f(x)$ of Eq.~(\ref{eq:2-16}) and of Ref.~\cite{BBKS}. We
also need $f_{10}$ and $f_{01}$ in the following calculations. They
are given by
\begin{align}
  \label{eq:3-33a}
  f_{10}(x)
  &= 
  -\frac{3x}{2}
  \left[
    \mathrm{erf}\left(\frac{1}{2}\sqrt{\frac{5}{2}}\,x\right) 
    + \mathrm{erf}\left(\sqrt{\frac{5}{2}}\,x\right) 
    \right]
  \nonumber\\
  & \quad - \frac{12\sqrt 2}{5\sqrt {5\pi}}
  \Biggl[
    e^{-5x^2/2}
 \!  -\! \left(\!1\!+\!\frac{15x^2}{8}\!\right)
      \left(\!1
      \!+\!\frac{15x^2}{16}
      \!\right) e^{-5x^2/8}
  \Biggr],
\\
  \label{eq:3-33b}
  f_{01}(x)
  &= 
  -\frac{\sqrt{7}}{5}
  \left[
    \mathrm{erf}\left(\frac{1}{2}\sqrt{\frac{5}{2}}\,x\right) 
    + \mathrm{erf}\left(\sqrt{\frac{5}{2}}\,x\right) 
    \right]
    \nonumber\\
  &\quad
  + \!\frac{\!9x\!}{\sqrt{70\pi}}\!
  \left[\!
    \frac{2}{15} e^{-5x^2/2} \!+\!
    \left(
       \frac{11}{5}\! +\! \frac{x^2}{4}\! + \!\frac{5x^4}{16}
     \right) e^{-5x^2/8}\!
  \right].
\end{align}
Hence, the integrals of Eq.~(\ref{eq:3-30}) can be reduced to
combinations of one-dimensional integrals.

\subsection{\label{subsec:Evaluation}
Evaluation of peak clustering with 1D FFTs
}

We have concrete expressions for the renormalized bias functions $c_n$
of peaks up to third order. Substituting
Eqs.~(\ref{eq:3-25a})--(\ref{eq:3-25c}) into Eq.~(\ref{eq:3-23}), we
obtained an analytic expression for the power spectrum of peaks.
However, the expression contains high-dimensional integrations which
is computationally expensive. For the third-order term, naively we
need to evaluate a nine-dimensional integrations, while statistical
isotropy reduces the integral down to seven dimensions, which is still
computationally expensive.

Recently, this type of integrations was found to be expressible in
terms of combinations of one-dimensional integrals in configuration
space \cite{SVM16,SV16,MFHB16,FBMH17}. The key technique is to expand
the Dirac delta function in Eq.~(\ref{eq:3-23-1}) into plane waves and
then expand them as a sum of spherical waves using spherical
Harmonics. The scalar products of the integrands, $[c_n]^2$ are also
expanded in spherical Harmonics, and finally, we can perform all the
angular integrals of Eq.~(\ref{eq:3-23}).

For $n=2$, we have a formula \cite{SVM16}
\begin{multline}
  \label{eq:3-34}
  \int_{\bm{k}_{12}=\bm{k}} {k_1}^{n_1} {k_2}^{n_2}
  \left(\hat{\bm{k}}_1\cdot\hat{\bm{k}}_2\right)^l
  P_\mathrm{s}(k_1) P_\mathrm{s}(k_2)
  =
  4\pi \int_0^\infty r^2dr\,j_0(kr)
  \\ \times
  \sum_{L=0}^l (-1)^L(2L+1)\,\alpha_{lL}\,
  \xi^{(n_1)}_L(r)\,\xi^{(n_2)}_L(r),
\end{multline}
where
\begin{equation}
  \label{eq:3-35}
  \xi_L^{(n)}(r) \equiv
  \int \frac{k^2dk}{2\pi^2} k^n j_L(kr) P(k) W^2(kR)
\end{equation}
are generalized correlation functions of the smoothed density field,
and
\begin{align}
  \label{eq:3-36}
  \alpha_{lL}
  &\equiv
    \frac{1}{2} \int_{-1}^1 d\mu\,\mu^l \mathsf{P}_{L}(\mu)
    \nonumber\\
  &= 
  \begin{cases}
    \displaystyle
    \frac{l!}{2^{(l-L)/2}[(l-L)/2]!\,(l+L+1)!!}
    & \left(
      \begin{matrix} l\geq L,\\ l+L=\mathrm{even}
        \end{matrix}
        \right), \\
    0 & (\mathrm{otherwise}),
  \end{cases}
\end{align}
with Legendre polynomials $\mathsf{P}_{L}(\mu) = (2^n
n!)^{-1}(d/dx)^n[(x^2-1)]$. 
For $n=3$, we have a formula \cite{SV16}
\begin{multline}
  \label{eq:3-37}
  \!\!\!\!\!
  \int_{\bm{k}_{123}=\bm{k}} \!\!\!\!\!\!\!\!\!\!{k_1}^{n_1} {k_2}^{n_2} {k_3}^{n_3}
  \left(\hat{\bm{k}}_2\!\cdot\!\hat{\bm{k}}_3\!\right)^{l_1}
  \left(\hat{\bm{k}}_3\!\cdot\!\hat{\bm{k}}_1\!\right)^{l_2}
  \left(\hat{\bm{k}}_1\!\cdot\!\hat{\bm{k}}_2\!\right)^{l_3}
 \! P_\mathrm{s}(k_1)  P_\mathrm{s}(k_2) P_\mathrm{s}(k_3)
  \\
  =
  4\pi \!\!\int_0^\infty\! \!\!\!\!\! r^2dr\,j_0(kr)\!\!
  \sum_{L_1=0}^{l_2+l_3}
  \sum_{L_2=0}^{l_3+l_1}
  \sum_{L_3=0}^{l_1+l_2}\!
  \mathcal{M}^{L_1L_2L_3}_{l_1l_2l_3}
  \xi^{(n_1)}_{L_1}(r)\,
  \xi^{(n_2)}_{L_2}(r)\,
  \xi^{(n_3)}_{L_3}(r),
\end{multline}
where
\begin{equation}
  \label{eq:3-38}
  \mathcal{M}^{L_1L_2L_3}_{l_1l_2l_3} \equiv
  \sum_{l_1'=0}^{l_1} \sum_{l_2'=0}^{l_2}
  \sum_{l_3'=0}^{l_3}
  \alpha_{l_1l_1'}\alpha_{l_2l_2'}\alpha_{l_3l_3'}
  \left[
    \begin{array}{ccc}
      L_1 & L_2 & L_3 \\
      l_1' & l_2' & l_3' \\
    \end{array}
  \right],
\end{equation}
and
\begin{multline}
  \label{eq:3-39}
  \left[
    \begin{array}{ccc}
      L_1 & L_2 & L_3 \\
      l_1 & l_2 & l_3 \\
    \end{array}
  \right] \equiv
  i^{L_1+L_2+L_3}(-1)^{l_1+l_2+l_3}
  (2L_1\!+\!1)(2L_2\!+\!1)
  \\ \times
  (2L_3\!+\!1)
  (2l_1\!+\!1)(2l_2\!+\!1)(2l_3\!+\!1)
  \left(
    \begin{array}{ccc}
      L_1 & L_2 & L_3 \\
      0 & 0 & 0 \\
    \end{array}
  \right)
  \\ \times
  \left(
    \begin{array}{ccc}
      L_1 & l_2 & l_3 \\
      0 & 0 & 0 \\
    \end{array}
  \right)
  \left(
    \begin{array}{ccc}
      L_2 & l_1 & l_3 \\
      0 & 0 & 0 \\
    \end{array}
  \right)
  \left(
    \begin{array}{ccc}
      L_3 & l_1 & l_2 \\
      0 & 0 & 0 \\
    \end{array}
  \right)
  \left\{
    \begin{array}{ccc}
      L_1 & L_2 & L_3 \\
      l_1 & l_2 & l_3 \\
    \end{array}
  \right\}
  \nonumber
\end{multline}
is a rescaled 6$j$-symbol.

Applying the above formulas to Eq.~(\ref{eq:3-23}) up to third order
yields a large number of terms, which is conveniently
manipulated by a software package such as \textsl{Mathematica}. The
sign convention of 6$j$-symbols in this paper agrees with that of
\textsl{Mathematica}. The results are given in a form
\begin{equation}
  \label{eq:3-40}
  P_\mathrm{pk}(k) = 4\pi\int_0^\infty r^2dr\,j_0(kr)
  \xi_\mathrm{pk}(r),
\end{equation}
where
\begin{equation}
  \label{eq:3-41}
  \xi_\mathrm{pk}(r) =
  \sum_{N=1}^\infty \frac{\xi_\mathrm{pk}^{(N)}(r)}{N!}
\end{equation}
is the correlation function of peaks, and $\xi_\mathrm{pk}^{(N)}(r)$
is given by a sum of $N$ products of generalized correlation
functions, $\xi^{(n)}_L(r)$. Explicit expressions of
$\xi_\mathrm{pk}^{(N)}(r)$ for $N=1,2,3$ are given in
Appendix~\ref{app:third}.

The integrals of generalized correlation functions are one-dimensional
Hankel transforms, which can be efficiently evaluated with a fast
Fourier transform (FFT) using \textsl{FFTLog} \cite{Ham00}. Once the
correlation function of peaks is evaluated, the power spectrum can be
obtained by a Hankel transform again. This procedure is very efficient
and robust.

\subsection{\label{subsec:NonGaussianity}
Effects of non-Gaussianity
}

In this subsection, we briefly consider the effect of non-Gaussianities
in the underlying field. The effect of non-Gaussianities can be
evaluated within the iPT formalism \cite{Mat11,Mat14}. Since we are
interested in the initial conditions, we ignore the components of
gravitational evolution in the iPT formalism. The non-Gaussian
contributions to the power spectrum is given by (see, e.g., Eq.~[13]
of Ref.~\cite{Mat12} with substitutions $\Gamma_X^{(n)} = c_n$)
\begin{equation}
  \label{eq:3-50}
  P^\mathrm{NG}_\mathrm{pk}(k) = c_1(k)
  \int_{\bm{k}_{12}=\bm{k}} c_2(\bm{k}_1,\bm{k}_2)
  B_\mathrm{s}(k,k_1,k_2),
\end{equation}
where
$B_\mathrm{s}(k_1,k_2,k_3) \equiv B(k_1,k_2,k_3)W(k_1R)W(k_2R)W(k_3R)$
is the bispectrum of the smoothed density field, and the bispectrum of
the underlying field is defined by
\begin{equation}
  \label{eq:3-51}
  \left\langle
    \tilde{\delta}(\bm{k}_1)
    \tilde{\delta}(\bm{k}_2)
    \tilde{\delta}(\bm{k}_3)
  \right\rangle
  =(2\pi)^3 \delta_\mathrm{D}^3(\bm{k}_1+\bm{k}_2+\bm{k}_3)
  B(k_1,k_2,k_3).
\end{equation}

The angular integrations in Eq.~(\ref{eq:3-50}) can be analytically
performed along the same line of calculations as in
Sec.~\ref{subsec:Evaluation}. Specifically, one can derive the formula
\begin{multline}
  \label{eq:3-52}
  \int_{\bm{k}_{12}=\bm{k}} {k_1}^{n_1} {k_2}^{n_2}
  \left(\hat{\bm{k}}_1\cdot\hat{\bm{k}}_2\right)^l
  B_\mathrm{s}(k,k_1,k_2)
\\ =
  4\pi \int_0^\infty r^2dr\,j_0(kr)
  \sum_{L=0}^l (-1)^L(2L+1)\,\alpha_{lL}
  \\ \times
  \int \frac{{k_1}^2dk_1}{2\pi^2} \frac{{k_2}^2dk_2}{2\pi^2}
  {k_1}^{n_1} {k_2}^{n_2}
  j_L(k_1r) j_L(k_2r) B_\mathrm{s}(k,k_1,k_2).
\end{multline}
The last two integrals over $k_1$ and $k_2$ are separated when the
bispectrum is given by a sum of factorized products of functions with
arguments $k_1$ and $k_2$.

For example, the initial bispectrum of curvature perturbations $\zeta$
in the presence of local-type non-Gaussianities is given by
\begin{equation}
  \label{eq:3-53}
  B_\zeta(k_1,k_2,k_3) =
  \frac{6}{5} f_\mathrm{NL}
  \left[
    P_\zeta(k_1) P_\zeta(k_2) + \mathrm{cyc.}
  \right],
\end{equation}
where $P_\zeta(k)$ is the power spectrum of the initial curvature
perturbations, and $f_\mathrm{NL}$ is a non-Gaussianity parameter. In
this case, the bispectrum $B_\mathrm{s}(k_1,k_2,k_3)$ is also given by
a sum of products of the power spectrum $P_\mathrm{s}(k)$ with
$k$-dependent coefficients. Here, the last two-dimensional integral in
Eq.~(\ref{eq:3-52}) is separated as a product of one-dimensional
integrals which can be evaluated using \textsl{FFTLog}. In
Appendix~\ref{app:localNG}, Eq.~(\ref{eq:3-50}) for the local-type
non-Gaussianity is explicitly derived.

\subsection{\label{subsec:CalcSummary}
Summary of the obtained predictions for peak clustering
}

We have derived all the necessary formulas to calculate the
correlation function and the power spectrum of peaks up to third order
in the generalized correlation functions $\xi^{(n)}_l(r)$ or the power
spectrum $P_\mathrm{s}(k)$ of the underlying (Gaussian) density field.
The correlation function of peaks is given by Eq.~(\ref{eq:3-41}),
where $\xi^{(N)}_\mathrm{pk}(r)$ is given by
Eqs.~(\ref{eq:d-1})--(\ref{eq:d-3}) for $N=1,2,3$. The bias
coefficients in the last equations are calculated using
Eq.~(\ref{eq:3-30}) and (\ref{eq:3-26}) with Eqs.~(\ref{eq:3-31a}),
(\ref{eq:b-1}), (\ref{eq:2-15-1}), (\ref{eq:2-16}), (\ref{eq:3-33a})
and (\ref{eq:3-33b}). In practice, all the necessary integrations can
be reduced to combinations of one-dimensional integrals as described
above. The formula for $N=3$ is too tedious to be manually handled.
Practically, we use the software package \textsl{Mathematica} to
derive, manipulate and numerically evaluate those tedious terms. The
Mathematica code of the FFTLog package by
Hamilton\footnote{\texttt{http://jila.colorado.edu/\~{}ajsh/FFTLog/}}
makes it easy to calculate everything in a Mathematica notebook. The
power spectrum of peaks, $P_\mathrm{pk}(k)$, is eventually given by
Eq.~(\ref{eq:3-40}).

\section{\label{sec:SampleCalc}
A Sample calculation
}

The main purpose of this paper is to provide analytic expressions for
the correlation function and power spectrum of peaks. Below, some
numerical calculations are presented as an example of applications. We
consider the underlying power spectrum after the decoupling epoch with
cold dark matter plus baryons. The motivation for this example is the
application to the biasing problem in the context of structure
formation, as the derived correlation function and power spectrum
provides us with the clustering of peaks in Lagrangian space.

\subsection{A MCMC estimate of the full peak correlation function}
\label{sec:MCMC}

In order to assess the validity of our bias expansion approach, we
compare to a full numerical integration of the peak correlation
function obtained by an MCMC method implemented in {\sl Mathematica}.
This strategy was already used to compute the correlation function of
peaks in 1D in \cite{BCDP16} and in 3D in \cite{CPP18}. In practice,
random numbers of dimension 14 are drawn from the conditional
probability that $(\nu,\zeta_{ij})$ at position $\bm{x}_{1}$ and
$(\nu,\zeta_{ij})$ at position $\bm{x}_{2}$ satisfy the Gaussian
distribution given that $\eta_i=0$. For each draw ${}^{(k)}$, we keep
the sample only if heights are above the threshold $\nu_\mathrm{c}$
and curvatures (eigenvalues of $\bm{\zeta}$) are negative and we
evaluate
$ \mathrm{det}[\zeta^{(k)}_{ij}(\bm{x}_{1})]
\mathrm{det}[\zeta^{(k)}_{ij}(\bm{x}_{2})]$.
\begin{multline}
  \left\langle
    N_\mathrm{pk}(\nu_\mathrm{c},\bm{x}_{1})
    N_\mathrm{pk}(\nu_\mathrm{c},\bm{x}_{2})
  \right\rangle\\
  \approx \frac{ \!\mathcal{P}[\bm{\eta}(\bm{x}_{1})
    \!=\!\bm{\eta}(\bm{x}_{2})\!=\!\bm{0}]}{N}
  \!\sum_{k\in {\cal S}} \!
  \mathrm{det}\left[\zeta^{\!(k)}_{ij} (\bm{x}_{1})\right]
  \mathrm{det}\left[\zeta^{\!(k)}_{ij}(\bm{x}_{2})\right],
    \nonumber
\end{multline}
where $N$ is the total number of draws, and $\cal S$ is the subset of
the indices of draws satisfying the constraints on the eigenvalues and
heights. The same procedure can be applied to evaluate the denominator
appearing in the expression of the peak correlation function. In
practice, the constraint on the peak height and the dimensionality of
the problem makes this calculation very expensive. However, because
the algorithm is embarrassingly parallel, we use a local cluster to
perform the calculation in a reasonable amount of human time.

In this paper, for each separation, we performed 7 estimates of the
correlation function (in order to get an estimate of the error bars by
measuring the error on the mean from those 7 estimates). For each of
them, we drew 5 billions times 12 random numbers to perform the MCMC,
an evaluation that we parallelised on 64 cores. On our cluster, each
of the 7 estimates took on average 30 hours (with some variability).

\subsection{Comparison between MCMC integrations and bias expansion}

\begin{figure*}
\begin{center}
\includegraphics[height=13.3pc]{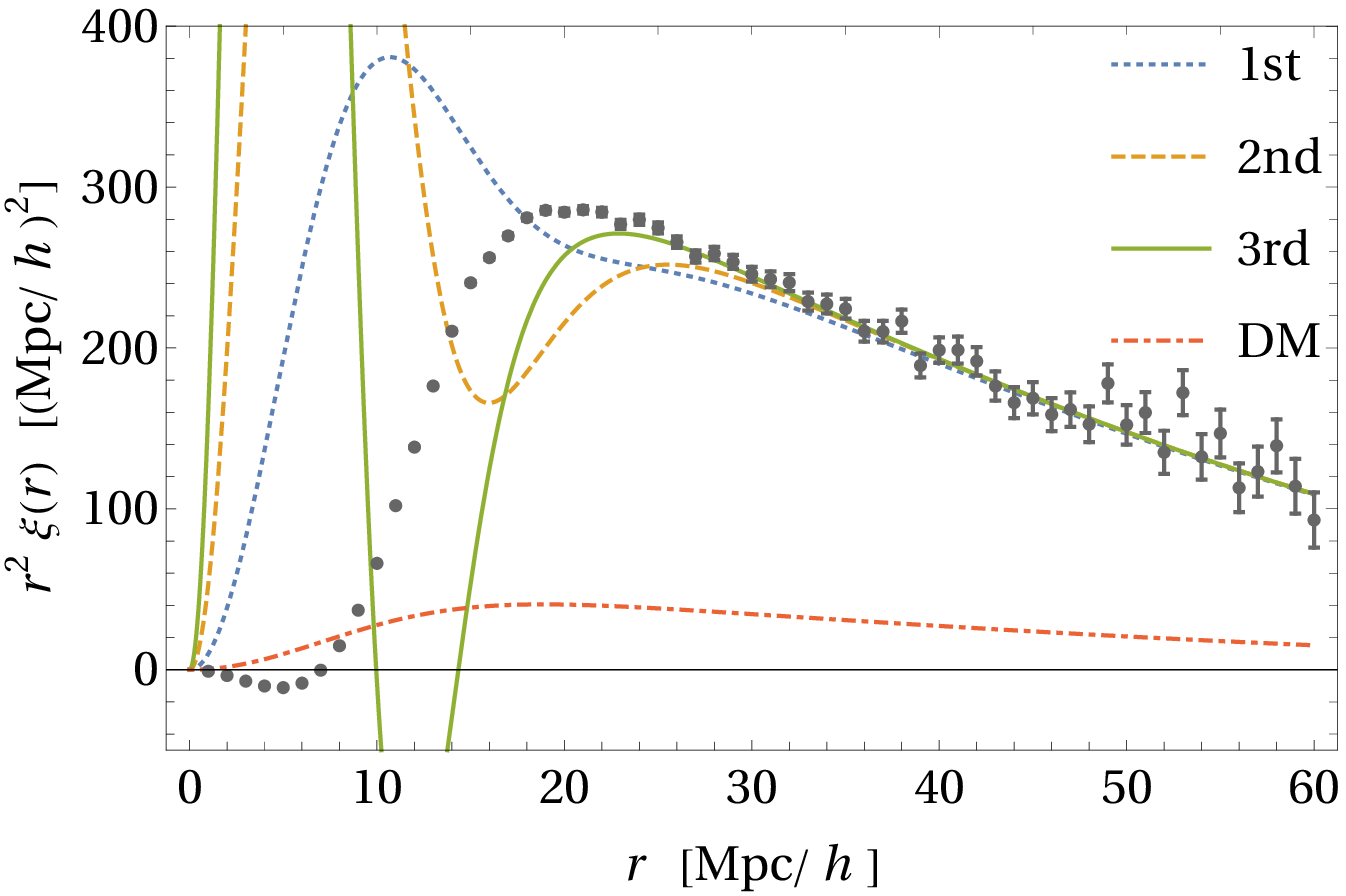}
\hspace{1pc}
\includegraphics[height=13.3pc]{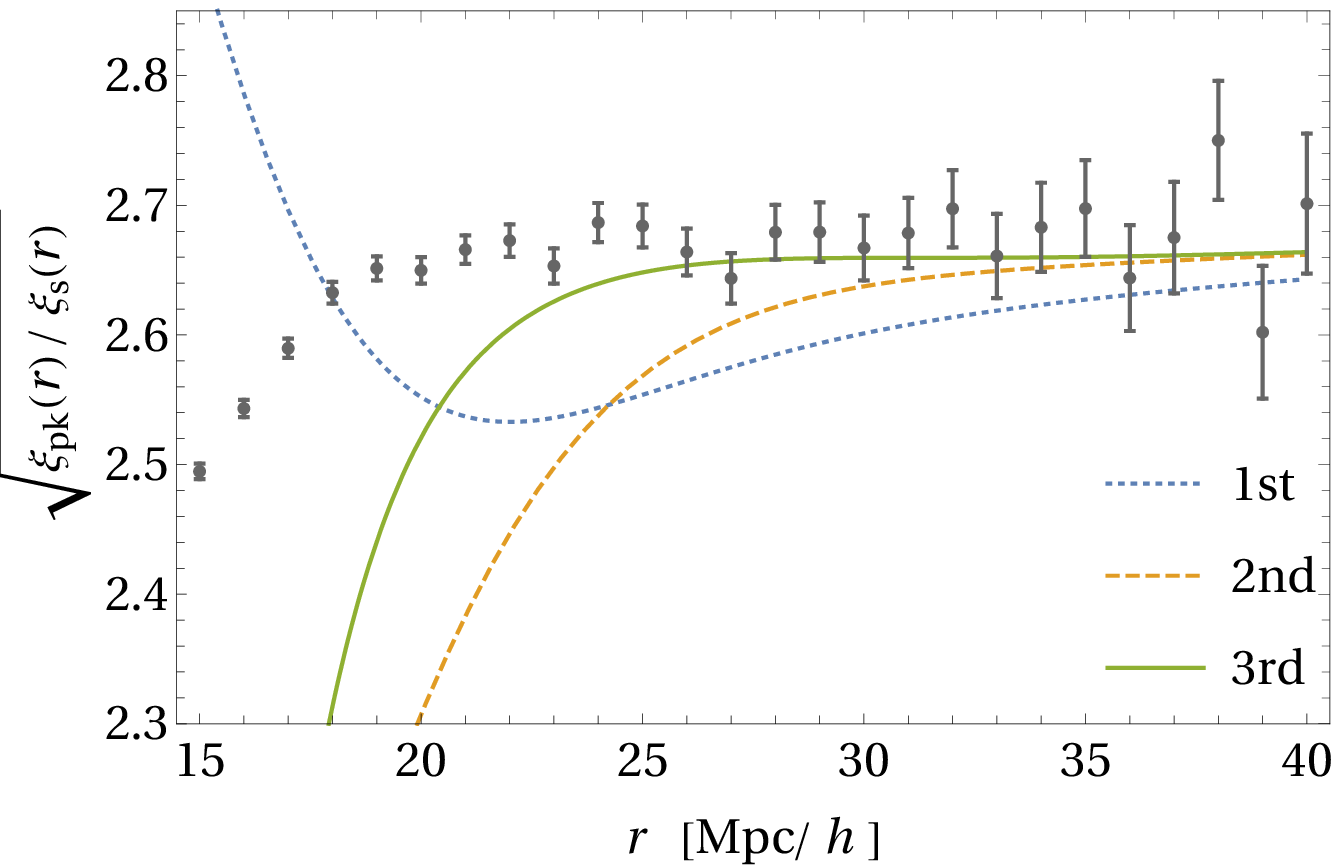}
\caption{\label{fig:cf_cdm} Correlation functions for matter and peaks
  of height $\nuc=2.5$ with a flat $\Lambda$CDM model. {\em Left
    panel}: Correlation functions of matter (blue) and of peaks with
  1st-order (orange), 2nd-order (green), and 3rd-order (red)
  approximations. The points with error bars represent the results of
  numerical integrations of peak correlation function. {\em Right
    panel}: The scale-dependent bias of peaks with corresponding
  approximations.}
\end{center}
\end{figure*}
In the following calculations, the power spectrum of the underlying
density field is calculated by a Boltzmann code \textsl{CLASS}
\cite{class11,CLASS} with a flat $\Lambda$CDM model and cosmological
parameters $h=0.6732$, $\Omega_{\mathrm{b}0}h^2=0.02238$,
$\Omega_{\mathrm{cdm}}h^2=0.1201$, $n_\mathrm{s}=0.9660$,
$\sigma_8=0.8120$ (Planck 2018 \cite{Planck2018}). We apply a
smoothing radius of $R=5\,h^{-1}\mathrm{Mpc}$ for the underlying mass
density field, and apply a peak threshold of $\nuc=2.5$.
Fig.~\ref{fig:cf_cdm} shows the correlation function of peaks given by
Eq.~(\ref{eq:3-41}), with approximations up to 1st, 2nd and 3rd
orders. The correlation function of the underlying mass density field
is also plotted. The points with error bars are the results of a
14-dimensional Monte-Carlo integration of the full expression of the
peak correlation function as described in Sec.~\ref{sec:MCMC} .

In the left figure, the correlation functions are multiplied by $r^2$.
The three approximations are almost identical on large scales
$r\gtrsim 40\,h^{-1}\mathrm{Mpc}$, which means that the 1st-order
approximation is already accurate enough to describe the correlation
function on these scales. The three approximations then deviate from
each other on smaller scales, $r \lesssim 40\,h^{-1}\mathrm{Mpc}$. A
remarkable structure of the higher-order approximations is the
decrease of the correlation function on small scales which is seen
around $r\sim 20\,h^{-1}\mathrm{Mpc}$. This structure is explained by
the exclusion effect of peaks \cite{BCDP16,CPP18}: the peaks cannot be
too close to each other in the smoothed density field in particular
because of the constraint on the sign of the curvatures. Hence, the
correlation function of peaks should be negative around the scales of
the smoothing radius. In the Monte-Carlo integration, this exclusion
effect is accurately seen as expected. In the perturbative
approximations, the exclusion effect is not sufficiently seen in the
first-order approximation. In the third-order approximation, the
correlation function actually becomes negative on small scales.
However, higher-order contributions becomes more important on
sufficiently small scales, and behaviors on small scales in our
approximations do not correspond to the true correlation function of
peaks, highlighting the rather poor convergence of the perturbative
bias expansion on small scales which can not capture the
non-perturbative exclusion zone.

However, the perturbative method and the Monte-Carlo approach are
complementary. The small-scale behaviors, including the exclusion
effect, are accurately captured by a Monte-Carlo method, while the
large-scale behaviors are more accurately and efficiently evaluated
by perturbative methods. The computational time of the perturbative
method is of the order of a few seconds for this figure, while that of
the Monte-Carlo method is of the order of a couple of days on 7 nodes
with 64 cores each. The perturbative method is suited for evaluating
large-scale correlations, while the Monte-Carlo method is better
suited for evaluating small-scale correlations and exclusion effects.

To precisely see differences among approximations of various orders,
the scale-dependent bias,
$[\xi_\mathrm{pk}(r)/\xi_\mathrm{s}(r)]^{1/2}$, is plotted in the
right-hand panel of Fig.~\ref{fig:cf_cdm}, where
$\xi_\mathrm{s}(r) = \xi^{(0)}_0(r)$ is the correlation function of
the underlying mass density field. This figure shows percent-level
differences among the different approximations. The differences
between first-order and higher-order approximations are within a
percent for $r \gtrsim 40\,h^{-1}\mathrm{Mpc}$. The difference between
second-order and third-order approximations are within a percent for
$r \gtrsim 30\,h^{-1}\mathrm{Mpc}$. Therefore, we can reliably apply
second-order approximation in the last range of scales. Comparing with
the results of Monte-Carlo integration, the third-order approximation
looks accurate at the percent-level for
$r \gtrsim 25\,h^{-1}\mathrm{Mpc}$.

\begin{figure*}
\begin{center}
  \includegraphics[height=12.5pc]{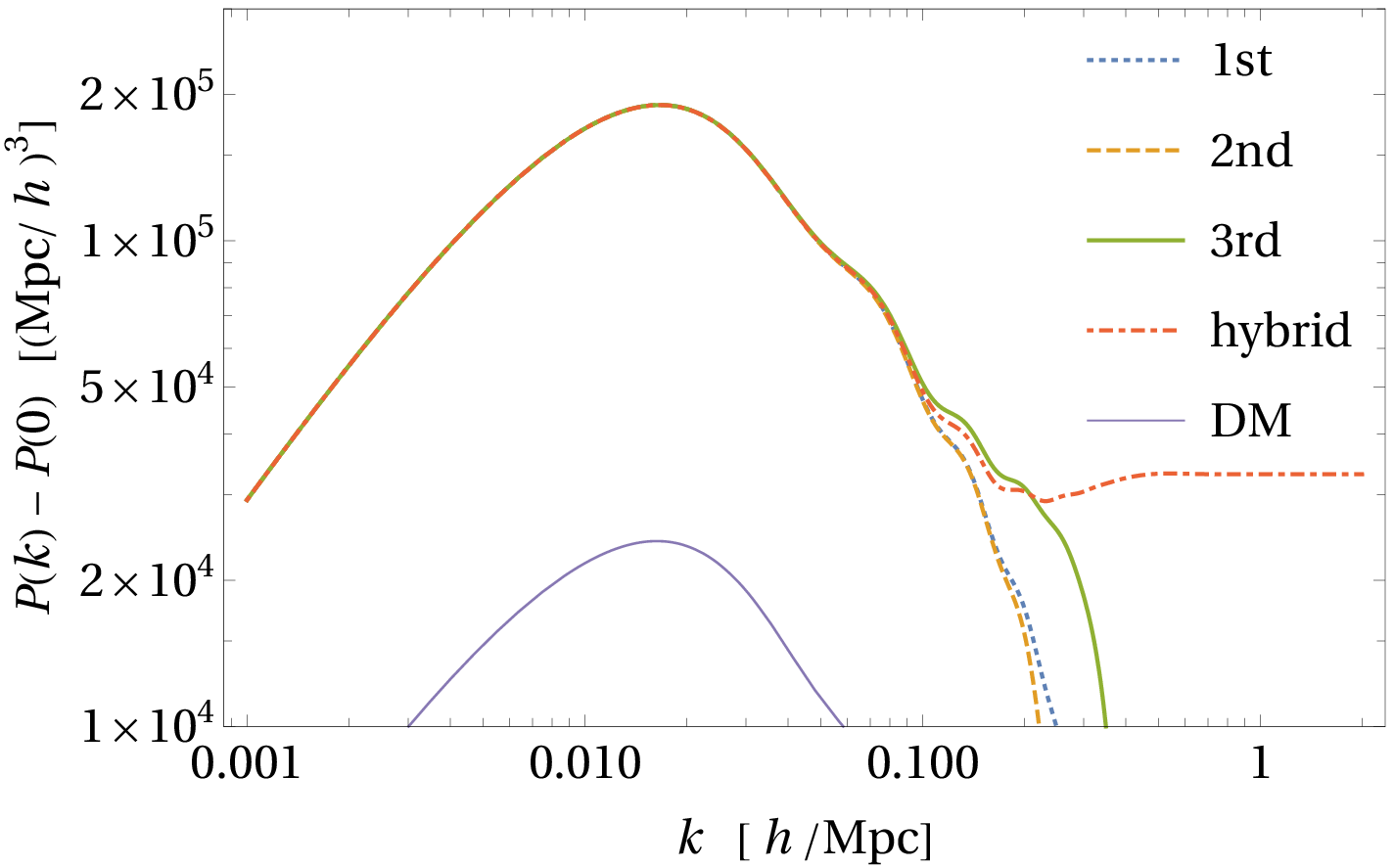}
  \hspace{1pc}
  \includegraphics[height=12.5pc]{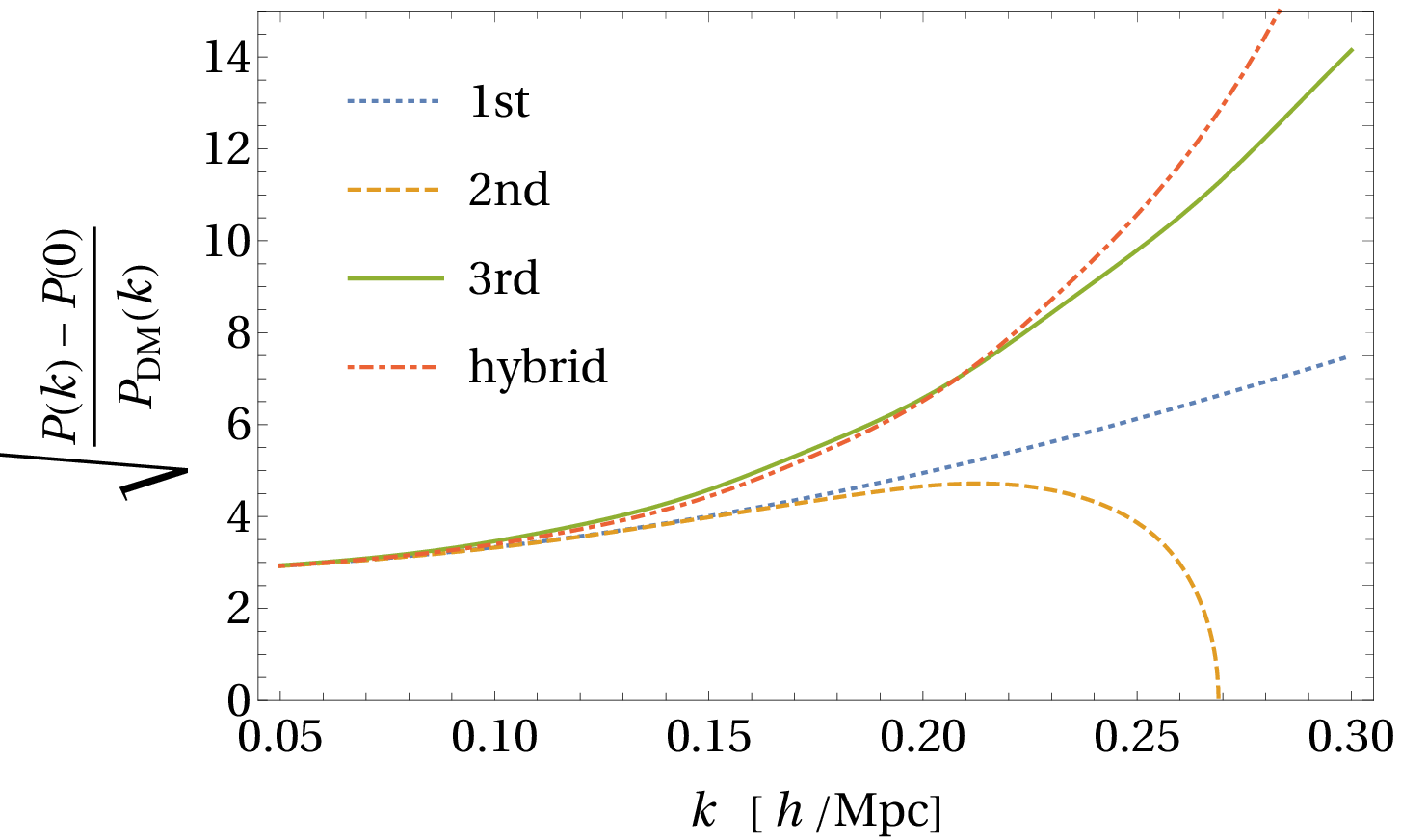}
  \caption{\label{fig:ps_cdm} Power spectra in the linear density
    field with a flat $\Lambda$CDM model. The zero-lag value
    $P(k\rightarrow 0)$ is subtracted from each power spectrum. {\em
      Left panel}: Power spectra of matter (violet) and of peaks with
    1st-order (blue), 2nd-order (orange), 3rd-order (green)
    approximations, and the ``hybrid'' data (red) are plotted. {\em
      Right panel}: The scale-dependent bias of the power spectra
    (with zero-lag values subtracted off) with corresponding
    approximations. }
\end{center}
\end{figure*}

Next we consider the power spectrum of peaks by Fourier transforming
the correlation function obtained above. The behavior of the
correlation function below the scales of the exclusion zone
($\lesssim 20\,h^{-1}\mathrm{Mpc}$) non-trivially affects the power
spectrum even on large scales ($k \rightarrow 0$), engendering a
sub-Poissonian behavior of peaks that was already discussed in
Refs.~\cite{Bal13,BCDP16} in 1D. This effect of exclusion zone has
highly non-perturbative nature and cannot be captured by the large-separation
expansions. In this paper, we just remove the effect of
exclusion zone by subtracting off the zero-lag value
$P(k\rightarrow 0)$ from the power spectra, and the results are
given in Fig.~\ref{fig:ps_cdm}.

In this figure, we also plot the power spectrum from the ``hybrid''
correlation function, which is a composition of the interpolated MCMC
results for $r<30\,h^{-1}\mathrm{Mpc}$ and 3rd-order approximation for
$r>30\,h^{-1}\mathrm{Mpc}$. Neglecting the constant components from
the effect of exclusion zone, the power spectrum of the
large-separation expansion gives a good prediction for
$k\lesssim 0.1$--$0.2\,h\mathrm{Mpc}^{-1}$. However, the
large-separation expansions gives positive value of
$P(k\rightarrow 0)$ while the hybrid data including the exclusion
effects gives a negative value of the same quantity (which is why
removing this negative constant to get the red dot-dashed line gives a
positive plateau on small scales). A more detailed investigation of
the effect of the exclusion zone in 3D is beyond the scope of this
paper which focuses on large-separation expansions, and will be
investigated thoroughly in a dedicated paper.

\section{\label{sec:Concl}
Conclusions
}

In this paper, we revisit the problem of estimating peak clustering
within a large-separation expansion. We derive a
set of formulas to evaluate the correlation function of peaks in
Gaussian random fields in general. The renormalized bias functions of
peaks are derived from the definition up to third order, and the
resultant correlation function of peaks are represented by an analytic
form which can be evaluated with 1D FFT only. Thus the numerical
evaluations of the formula is very fast.

The result of the derived formula are compared with a Monte-Carlo 14D
integrations of the exact expression of the peak correlation function.
The results are consistent with each other on large scales, where the
large-separation expansion is sufficiently accurate
($r \gtrsim 25\,h^{-1}\mathrm{Mpc}$). On small scales, the Monte-Carlo
method is able to provide an accurate estimate including the exclusion
effect which is essentially non-perturbative and cannot be captured by
our bias expansion. Therefore, both methods are quite complementary to
each other.

In connection to the structure formation, the correlation of peaks in
this paper are evaluated in Lagrangian space. Dynamical evolutions of
the peak positions should be taken into account for the predictions in
Eulerian space. The iPT \cite{Mat11,Mat12,Mat14,MD16} offers a
systematic method to perturbatively evaluate the clustering in
Eulerian space. Including the dynamical evolutions by iPT, the results
of higher-order perturbation theory should also be reduced to
expressions with 1D FFT, using the technique of
Ref.~\cite{SVM16,SV16,MFHB16,FBMH17}. Deriving the concrete
expressions in Eulerian space using the iPT up to third order will be
addressed in future work.

\begin{acknowledgments}
  The authors thank Tobias Baldauf, Vincent Desjacques, Kazunori
  Kohri, Christophe Pichon, Dmitri Pogosian, Takahiro Terada for
  fruitful discussions together with the Yukawa Institute for
  theoretical physics and the organizers of the workshop PTchat@Kyoto
  during which part of this work was initiated. This work was
  supported by JSPS KAKENHI Grants No.~JP16H03977, No.~JP19K03835
  (TM). SC's work is partially supported by the SPHERES grant
  ANR-18-CE31-0009 of the French {\sl Agence Nationale de la
    Recherche} and by Fondation MERAC. This work has made use of the
  Horizon Cluster hosted by Institut d'Astrophysique de Paris. We
  thank Stephane Rouberol for running smoothly this cluster for us.
\end{acknowledgments}

\onecolumngrid
\appendix

\section{\label{app:cthree}
A direct derivation of the renormalized bias functions of peaks
}

In this Appendix, the explicit form of renormalized bias functions of
peaks up to third order, Eqs.~(\ref{eq:3-25a})--(\ref{eq:3-25c}), are
derived from the original definition of the functions. The first and
second order results are already given in Ref.~\cite{MD16}. This
Appendix is a straightforward generalization of the last work to the
third order. In this paper, the window function $W(kR)$ is included in
the power spectrum $P_\mathrm{s}(k)$ in Eq.~(\ref{eq:3-23}), while it
is included in the renormalized bias functions $c_n$ in
Ref.~\cite{MD16}.

The distribution function $P(\bm{y})$ is given by Eq.~(\ref{eq:2-11}).
The Fourier transform of the variables $y_\alpha(\bm{x})$ has a form,
\begin{equation}
  \label{eq:a-1}
  \tilde{y}_\alpha(\bm{k})
  = U_\alpha(\bm{k}) \tilde{\delta}_\mathrm{s}(\bm{k})
\end{equation}
where the functions $U_\alpha(\bm{k})$ are given by
\begin{equation}
  \label{eq:a-2}
  \left[ U_\alpha(\bm{k}) \right] =
  \left(
    \frac{1}{\sigma_0}, \frac{ik_1}{\sigma_1}, \frac{ik_2}{\sigma_1},
    \frac{ik_3}{\sigma_1}, \frac{-{k_1}^2}{\sigma_2},
    \frac{-{k_2}^2}{\sigma_2}, \frac{-{k_3}^2}{\sigma_2},
    \frac{-k_1k_2}{\sigma_2}, \frac{-k_2k_3}{\sigma_2},
    \frac{-k_1k_3}{\sigma_2}\right). 
\end{equation}
We define a differential operator 
\begin{equation}
  \label{eq:a-3}
  \mathcal{D}(\bm{k}) \equiv
  \sum_\alpha U_\alpha(\bm{k})
  \frac{\partial}{\partial y_\alpha} =
  \frac{1}{\sigma_0}\frac{\partial}{\partial\nu} +
  \frac{i}{\sigma_1}\bm{k}\cdot\frac{\partial}{\partial\bm{\eta}} -
  \frac{1}{\sigma_2} \sum_{i\leq j}
  k_i k_j \frac{\partial}{\partial\zeta_{ij}}.
\end{equation}
With this operator, the renormalized bias functions of
Eq.~(\ref{eq:3-24}) are given by
\begin{equation}
  \label{eq:a-4}
  c_n(\bm{k}_1,\ldots,\bm{k}_n) =
  \frac{1}{\bar{N}_\mathrm{pk}}
  \left\langle \mathcal{D}(\bm{k}_1) \cdots \mathcal{D}(\bm{k}_n)
    N_\mathrm{pk} \right\rangle
  = \frac{(-1)^n}{\bar{N}_\mathrm{pk}}
  \int d^{10}y\,N_\mathrm{pk}(\bm{y})
  \mathcal{D}(\bm{k}_1) \cdots \mathcal{D}(\bm{k}_n)
  \mathcal{P}(\bm{y}).
\end{equation}
The first equality in the above equation can be derived by Fourier
transforming Eq.~(\ref{eq:3-24}) with respect to $\bm{k}$, and partial
integrations are applied in the second equality.

Although independent set of variables $\zeta_{ij}$ are given for
$i\leq j$, it is useful to introduce a set of symmetric tensor
\begin{equation}
  \label{eq:a-5}
  \xi_{ij} \equiv
  \begin{cases}
    \zeta_{ij} & (i\leq j), \\
    \zeta_{ji} & (i > j).
  \end{cases}
\end{equation}
Any function of $\zeta_{ij}$ $(i\leq j)$ can be considered as a
function of $\xi_{ij}$. The differentiation with respect to variables
$\zeta_{ij}$ is given by
\begin{equation}
  \label{eq:a-6}
  \frac{\partial}{\partial\zeta_{ij}} = 
  \begin{cases}
    \frac{\partial}{\partial\xi_{ij}} & (i = j), \\
    \frac{\partial}{\partial\xi_{ij}}
      + \frac{\partial}{\partial\xi_{ji}} & (i < j).
  \end{cases}
\end{equation}
With these new variables, the differential
operator $\mathcal{D}(\bm{k})$ of Eq.~(\ref{eq:a-3}) reduces to 
\begin{equation}
  \label{eq:a-7}
  \mathcal{D}(\bm{k}) \equiv
  \sum_\alpha U_\alpha(\bm{k})
  \frac{\partial}{\partial y_\alpha} =
  \frac{1}{\sigma_0}\frac{\partial}{\partial\nu} +
  \frac{i}{\sigma_1}k_i\frac{\partial}{\partial\eta_i} -
  \frac{1}{\sigma_2}
  k_i k_j \frac{\partial}{\partial\xi_{ij}},
\end{equation}
where repeated indices $i,j$ are summed over. Rotationally invariant
quantities $\eta^2$, $J_1$, $J_2$, $J_3$ are just given by
Eqs.~(\ref{eq:2-12}) and (\ref{eq:2-13}) with replacements
$\zeta_{ij} \rightarrow \xi_{ij}$:
\begin{equation}
  \label{eq:a-8}
  \eta^2 \equiv \bm{\eta}\cdot\bm{\eta}, \quad
  J_1 \equiv -\xi_{ii}, \quad
  J_2 \equiv \frac{3}{2} \tilde{\xi}_{ij} \tilde{\xi}_{ji}, \quad
  J_3 \equiv \frac{9}{2} \tilde{\xi}_{ij} \tilde{\xi}_{jk}
  \tilde{\xi}_{ki},
\end{equation}
with
\begin{equation}
  \label{eq:a-9}
  \tilde{\xi}_{ij} \equiv \xi_{ij} + \frac{1}{3} \delta_{ij} J_1.
\end{equation}

To calculate the differentiations of the last expression of
Eq.~(\ref{eq:a-4}), the relations
\begin{equation}
  \label{eq:a-10}
  \frac{\partial(\eta^2)}{\partial\eta_i}= 2\eta_i, \quad
  \frac{\partial J_1}{\partial\xi_{ij}} = -\delta_{ij}, \quad
  \frac{\partial J_2}{\partial\xi_{ij}} = 3\tilde{\xi}_{ji}, \quad
  \frac{\partial\tilde{\xi}_{kl}}{\partial\xi_{ij}} =
  \delta_{ik}\delta_{jl} - \frac{1}{3} \delta_{ij}\delta_{kl}
\end{equation}
are useful. The distribution function $P(\bm{y})$ depends only on four
rotationally invariant variables $\nu$, $\eta^2$, $J_1$ and $J_2$ as
given in Eq.~(\ref{eq:2-11}). Using the above relations, the
first-order derivatives are given by
\begin{equation}
  \label{eq:a-11}
  \frac{\partial}{\partial\eta_i} \mathcal{P}
  = 2 \eta_i \frac{\partial}{\partial(\eta^2)} \mathcal{P}, \quad
  \frac{\partial}{\partial\xi_{ij}} \mathcal{P}
  = \left[
    - \delta_{ij} \frac{\partial}{\partial J_1} 
    + 3 \tilde{\xi}_{ji} \frac{\partial}{\partial J_2}
    \right] \mathcal{P},
\end{equation}
the second-order derivatives are given by
\begin{align}
  \label{eq:a-12a}
  \frac{\partial^2}{\partial\eta_i\partial\eta_j} \mathcal{P} &=
  2 \left[
     \delta_{ij} \frac{\partial}{\partial(\eta^2)}
  + 2 \eta_i \eta_j \frac{\partial^2}{\partial(\eta^2)^2}
  \right] \mathcal{P},
\\
  \label{eq:a-12b}
  \frac{\partial^2}{\partial\xi_{ij}\partial\xi_{kl}} \mathcal{P}
  &=
  \left[
  \delta_{ij} \delta_{kl} \frac{\partial^2}{\partial {J_1}^2}
  - 3 \left(
    \delta_{ij} \tilde{\xi}_{lk} + \delta_{kl}
    \tilde{\xi}_{ji} \right)
  \frac{\partial^2}{\partial J_1 \partial J_2} +
  9 \tilde{\xi}_{ji} \tilde{\xi}_{lk} 
  \frac{\partial^2}{\partial {J_2}^2}
  + \left(3 \delta_{il} \delta_{jk} - \delta_{ij} \delta_{kl}\right)
  \frac{\partial}{\partial J_2}
  \right] \mathcal{P},
\end{align}
and the third-order derivatives are given by
\begin{align}
  \label{eq:a-13a}
  \frac{\partial^3}{\partial\eta_i\partial\eta_j\partial\eta_k} \mathcal{P} &=
  4\left[
    \left(\delta_{ij}\eta_k + \delta_{jk}\eta_i + \delta_{ki}\eta_j\right)
      \frac{\partial^2}{\partial(\eta^2)^2} + 
    2 \eta_i \eta_j \eta_k \frac{\partial^3}{\partial(\eta^2)^3}
  \right] \mathcal{P},
\\
  \label{eq:a-13b}
  \frac{\partial^3}{\partial\xi_{ij}\partial\xi_{kl}\partial\xi_{mn}} \mathcal{P}
  &=
  \left[
  - \delta_{ij} \delta_{kl} \delta_{mn} \frac{\partial^3}{\partial {J_1}^3}
  + 3 \left(
    \delta_{ij} \delta_{kl} \tilde{\xi}_{nm} +
    \delta_{ij} \delta_{mn} \tilde{\xi}_{lk} +
    \delta_{kl} \delta_{mn} \tilde{\xi}_{ji}
    \right)
  \frac{\partial^3}{\partial {J_1}^2 \partial J_2} -
  9 \left(
    \delta_{ij} \tilde{\xi}_{lk}  \tilde{\xi}_{nm} +
    \delta_{kl} \tilde{\xi}_{ji}  \tilde{\xi}_{nm} +
    \delta_{mn} \tilde{\xi}_{ji}  \tilde{\xi}_{lk}
    \right)
  \frac{\partial^3}{\partial J_1 \partial {J_2}^2}
  \right.
\nonumber\\
  & \quad
  + 27 \tilde{\xi}_{ji} \tilde{\xi}_{lk} \tilde{\xi}_{nm} 
  \frac{\partial^3}{\partial {J_2}^3} +
  3 \left(
    \delta_{ij} \delta_{kl} \delta_{mn} -
    \delta_{ij} \delta_{kn} \delta_{lm} -
    \delta_{kl} \delta_{in} \delta_{jm} -
    \delta_{il} \delta_{jk} \delta_{mn}
    \right)
  \frac{\partial^2}{\partial J_1 \partial J_2}
\nonumber\\
  & \quad
  \left. +\,
  3 \left(
    3 \delta_{in} \delta_{jm} \tilde{\xi}_{lk}  +
    3 \delta_{kn} \delta_{lm} \tilde{\xi}_{ji}  +
    3 \delta_{il} \delta_{jk} \tilde{\xi}_{nm}  -
    \delta_{ij} \delta_{kl} \tilde{\xi}_{nm} -
    \delta_{ij} \delta_{mn} \tilde{\xi}_{lk} -
    \delta_{kl} \delta_{mn} \tilde{\xi}_{ji}
    \right)
  \frac{\partial^2}{\partial {J_2}^2}
  \right] \mathcal{P}.
\end{align}

In calculating Eq.~(\ref{eq:a-4}), one notices that the number density
$N_\mathrm{pk}$ and the distribution function $\mathcal{P}(\bm{y})$ only depend
on rotationally invariant variables. Thus we can first average over
the angular dependence in the product of operators $\mathcal{D}(\bm{k})$,
which appears only in the coefficients of
Eqs.~(\ref{eq:a-11})--(\ref{eq:a-13b}). Denoting the angular average
by $\langle\cdots\rangle_\Omega$, we have
\begin{align}
  \label{eq:a-14a}
  &
  \left\langle \eta_i \right\rangle_\Omega = 0, \quad
  \left\langle \eta_i \eta_j \right\rangle_\Omega = \frac{1}{3}
  \delta_{ij} \eta^2, \quad
  \left\langle \tilde{\xi}_{ij} \right\rangle_\Omega = 0, \quad
  \left\langle \tilde{\xi}_{ij} \tilde{\xi}_{kl}
  \right\rangle_\Omega =
  \frac{1}{15}
  \left(
      \delta_{ik} \delta_{jl} +  \delta_{il} \delta_{jk}
      - \frac{2}{3} \delta_{ij} \delta_{kl}
    \right) J_2,
\\
  \label{eq:a-14b}
  &
  \left\langle \tilde{\xi}_{ij} \tilde{\xi}_{kl} \tilde{\xi}_{mn}
  \right\rangle_\Omega =
  \frac{1}{315}
  \left[
    \frac{16}{3} \delta_{ij} \delta_{kl} \delta_{mn} -
    4 \left(
    \delta_{ij} \delta_{km} \delta_{ln} +
    \delta_{ij} \delta_{kn} \delta_{lm} +
    \delta_{kl} \delta_{im} \delta_{jn} +
    \delta_{kl} \delta_{in} \delta_{jm} +
    \delta_{mn} \delta_{ik} \delta_{jl} +
    \delta_{mn} \delta_{il} \delta_{jk}
    \right)
    \right.
\nonumber\\
  & \hspace{7pc}
  \left. +\,
    3 \left(
    \delta_{ik} \delta_{lm} \delta_{jn} +
    \delta_{jk} \delta_{lm} \delta_{in} +
    \delta_{il} \delta_{km} \delta_{jn} +
    \delta_{ik} \delta_{ln} \delta_{jm} +
    \delta_{jl} \delta_{km} \delta_{in} +
    \delta_{il} \delta_{kn} \delta_{jm} +
    \delta_{jk} \delta_{ln} \delta_{im} +
    \delta_{jl} \delta_{kn} \delta_{im}
    \right)
    \right] J_3.
\end{align}
The angular averages
can be taken for the operators in the integrand of Eq.~(\ref{eq:a-4}):\
\begin{equation}
  \label{eq:a-15}
  c_n(\bm{k}_1,\ldots,\bm{k}_n) =
  \frac{(-1)^n}{\bar{N}_\mathrm{pk}}
  \int d^{10}y\,N_\mathrm{pk}(\bm{y})
  \langle \mathcal{D}(\bm{k}_1)\cdots \mathcal{D}(\bm{k}_n)\rangle_\Omega
  \mathcal{P}(\bm{y}).
\end{equation}
Using the above equations, the results up to third order are given by
\begin{align}
  \label{eq:a-16a}
  \left\langle \mathcal{D}(\bm{k}) \right\rangle_\Omega \mathcal{P}
  &=
    \left(
    \frac{1}{\sigma_0} \frac{\partial}{\partial\nu} 
    + \frac{k^2}{\sigma_2} \frac{\partial}{\partial J_1}
    \right) \mathcal{P},
\\
  \label{eq:a-16b}
  \left\langle
    \mathcal{D}(\bm{k}_1) \mathcal{D}(\bm{k}_2)
  \right\rangle_\Omega \mathcal{P}
  &=
   \left\{
    \left(
      \frac{1}{\sigma_0} \frac{\partial}{\partial\nu} 
      + \frac{{k_1}^2}{\sigma_2} \frac{\partial}{\partial J_1}
    \right)
    \left(
      \frac{1}{\sigma_0} \frac{\partial}{\partial\nu} 
      + \frac{{k_2}^2}{\sigma_2} \frac{\partial}{\partial J_1}
    \right)
    \right.
\nonumber\\
  & \qquad
    \left.
   -
    \frac{2(\bm{k}_1\cdot\bm{k}_2)}{{\sigma_1}^2}
    \left[
      1 + \frac{2}{3} \eta^2 \frac{\partial}{\partial (\eta^2)}
    \right] \frac{\partial}{\partial(\eta^2)}
+    \frac{3(\bm{k}_1\cdot\bm{k}_2)^2 - {k_1}^2 {k_2}^2}{{\sigma_2}^2}
      \left[
        1 + \frac{2}{5} {J_2}^2 \frac{\partial}{\partial J_2}
      \right] \frac{\partial}{\partial J_2 }
    \right\} \mathcal{P},
\\
  \label{eq:a-16c}
  \left\langle
    \mathcal{D}(\bm{k}_1) \mathcal{D}(\bm{k}_2) \mathcal{D}(\bm{k}_3)
  \right\rangle_\Omega \mathcal{P}
  &=
    \left\{
    \left(
      \frac{1}{\sigma_0} \frac{\partial}{\partial\nu} 
      + \frac{{k_1}^2}{\sigma_2} \frac{\partial}{\partial J_1}
    \right)
    \left(
      \frac{1}{\sigma_0} \frac{\partial}{\partial\nu} 
      + \frac{{k_2}^2}{\sigma_2} \frac{\partial}{\partial J_1}
    \right)
    \left(
      \frac{1}{\sigma_0} \frac{\partial}{\partial\nu} 
      + \frac{{k_3}^2}{\sigma_2} \frac{\partial}{\partial J_1}
    \right)
    \right.
\nonumber\\
  & \qquad
   -
    \frac{2(\bm{k}_1\cdot\bm{k}_2)}{{\sigma_1}^2}
    \left(
      \frac{1}{\sigma_0} \frac{\partial}{\partial\nu} 
      + \frac{{k_3}^2}{\sigma_2} \frac{\partial}{\partial J_1}
    \right)
    \left[
      1 + \frac{2}{3} \eta^2 \frac{\partial}{\partial (\eta^2)}
    \right] \frac{\partial}{\partial(\eta^2)}
    + \mathrm{cyc.}
\nonumber\\
  & \qquad
    + \frac{3(\bm{k}_1\cdot\bm{k}_2)^2 - {k_1}^2 {k_2}^2}{{\sigma_2}^2}
    \left(
      \frac{1}{\sigma_0} \frac{\partial}{\partial\nu} 
      + \frac{{k_3}^2}{\sigma_2} \frac{\partial}{\partial J_1}
    \right)
      \left[
        1 + \frac{2}{5} {J_2}^2 \frac{\partial}{\partial J_2}
      \right] \frac{\partial}{\partial J_2}
    + \mathrm{cyc.}
\nonumber\\
  & \qquad
    \left.
    + \frac{72}{35}
    \left[
    (\bm{k}_1\cdot\bm{k}_2)(\bm{k}_2\cdot\bm{k}_3)(\bm{k}_3\cdot\bm{k}_1)
    - \frac{1}{3} {k_1}^2(\bm{k}_2\cdot\bm{k}_3) + \mathrm{cyc.}
    + \frac{2}{9} {k_1}^2 {k_2}^2 {k_3}^2 \right]
    J_3 \frac{\partial^3}{{\partial J_2}^3}
    \right\} \mathcal{P}.
\end{align}
Substituting identities
\begin{align}
  \label{eq:a-17}
    \left[
      1 + \frac{2}{3} \eta^2 \frac{\partial}{\partial (\eta^2)}
    \right] \frac{\partial}{\partial(\eta^2)} e^{-3\eta^2/2}
    &= - L^{(1/2)}_1\left(\frac{3}{2}\eta^2\right) e^{-3\eta^2/2},
\\
  \label{eq:a-18}
    \left[
      1 + \frac{2}{5} J_2 \frac{\partial}{\partial J_2}
    \right] \frac{\partial}{\partial J_2} e^{-5J_2/2}
    &= - L^{(3/2)}_1\left(\frac{5}{2} J_2\right) e^{-5J_2/2},
\\
  \label{eq:a-19}
  J_3 \frac{\partial^3}{{\partial J_2}^3} e^{-5J_2/2} &=
  \frac{75\sqrt{7}}{8} F_{01}(5J_2,J_3) e^{-5J_2/2}
\end{align}
into Eqs.~(\ref{eq:a-16a})--(\ref{eq:a-16c}) and (\ref{eq:a-15}), the
final results for the renormalized bias functions,
Eqs.~(\ref{eq:3-25a})--(\ref{eq:3-25c}) are obtained.

\section{\label{app:polynomials}
Orthogonal polynomials}

In this Appendix, we give the definitions of the orthogonal
polynomials of field variables to define the bias coefficients
$c_{ijqlm}$ in Eq.~(\ref{eq:3-27}). We closely follow the notation of
Refs.~\cite{GPP12,LMD16,Diz16} in this paper.

The first polynomials are the multivariate Hermite polynomials,
\begin{equation}
  \label{eq:b-1}
  H_{ij}(\nu, J_1)
  = \frac{1}{\mathcal{N}(\nu,J_1)} \left(-\frac{\partial}{\partial \nu}\right)^i
    \left(-\frac{\partial}{\partial J_1}\right)^j \mathcal{N}(\nu,J_1),
\end{equation}
where $\mathcal{N}(\nu,J_1)$ is the joint distribution function of $\nu$
and $J_1$ which is given by Eq.~(\ref{eq:2-15-1}). The multivariate
Hermite polynomials satisfy the orthonormality condition,
\begin{equation}
  \label{eq:b-3}
  \int_{-\infty}^\infty d\nu\,dJ_1\,\mathcal{N}(\nu,J_1)\,
  H_{ij}(\nu,J_1)\,H_{kl}(\nu,J_1)
  = i!\,j!\,\delta_{ik} \delta_{jl}.
\end{equation}
The second polynomials are the generalized Laguerre polynomials,
\begin{equation}
  \label{eq:b-4}
  L^{(\alpha)}_q(x) = \frac{x^{-\alpha} e^x}{q!} \frac{d^q}{dx^q}
  \left(x^{q+\alpha} e^{-x}\right).
\end{equation}
The Laguerre polynomials also satisfy an orthonormality condition,
\begin{equation}
  \label{eq:b-5}
  \int_0^\infty dx\,e^{-x} x^\alpha  L^{(\alpha)}_n(x) L^{(\alpha)}_{m}(x)
  = \frac{\Gamma(n + \alpha + 1)}{n!} \delta_{nm}.
\end{equation}
From this equation, we have
\begin{equation}
  \label{eq:b-6}
  \int_0^\infty d(3\eta^2)\,
  \chi_3^2(3\eta^2)
  L^{(\alpha)}_q\left(\frac{3}{2}\eta^2\right)
  L^{(\alpha)}_r\left(\frac{3}{2}\eta^2\right)
  = \frac{\Gamma(q + 3/2)}{q!\,\Gamma(3/2)} \delta_{qr},
\end{equation}
where
\begin{equation}
  \label{eq:b-6-1}
  \chi_k^2(x) \equiv \frac{x^{k/2-1} e^{-x/2}}{2^{k/2}\Gamma(k/2)}
\end{equation}
is a chi-square distribution with $k$ degrees of freedom, and
$\chi_3^2(3\eta^2)$ corresponds to the distribution function of
the variable $3\eta^2$.

The third polynomials are defined by\footnote{The
  function $F_{lm}$ here and the function $\tilde{F}_{lm}$ defined in
  Ref.~\cite{LMD16} are related by
  $\tilde{F}_{lm}=\sqrt{l!\Gamma(5/2)(2m+1)/\Gamma(l+5/2)}\,F_{lm}$.},
\begin{equation}
  \label{eq:b-7}
  F_{lm}(5J_2,J_3) = 
  (-1)^l\sqrt{
    \frac{\Gamma(l+5/2)}
    {\Gamma\left(l+3m+5/2\right)}}\,
    \left(\frac{5}{2}J_2\right)^{3m/2}
    L^{(3m+3/2)}_l\left(\frac{5}{2}J_2\right)\,
    \mathsf{P}_m\left(\frac{J_3}{{J_2}^{3/2}}\right),
\end{equation}
where
\begin{equation}
  \label{eq:b-8}
  \mathsf{P}_m(x) = \frac{1}{2^m m!} \frac{d^m}{dx^m} (x^2-1)^m
\end{equation}
are the Legendre polynomials, and satisfy the
orthonormality condition,
\begin{equation}
  \label{eq:b-9}
  \int_{-1}^1 dx\,e^{-x}\mathsf{P}_n(x)\,\mathsf{P}_m(x)
  = \frac{2}{2n+1} \delta_{nm}.
\end{equation}
The polynomials $F_{lm}$ satisfy orthonormality relations
\begin{equation}
  \label{eq:b-10}
  \int_0^\infty d(5J_2)\int_{-{J_2}^{3/2}}^{{J_2}^{3/2}}dJ_3\,
  p(5J_2,J_3)\,F_{lm}(5J_2,J_3)\,F_{l'm'}(5J_2,J_3)
  = \frac{\Gamma(l+5/2)}{\Gamma(5/2)\,l!\,(2m+1)} \delta_{ll'} \delta_{mm'},
\end{equation}
where
\begin{equation}
  \label{eq:b-11}
  p(5J_2,J_3) = \frac{\chi^2_5(5J_2)}{2 {J_2}^{3/2}}
    \Theta\left(1-{J_3}^2/{J_2}^3\right) =
    \frac{5^{3/2}}{6\sqrt{2\pi}} e^{-5J_2/2}
    \Theta\left(1-{J_3}^2/{J_2}^3\right)
\end{equation}
is the joint probability distribution function of the variables $5J_2$
and $J_3$. The above orthonormality relations and Eq.~(\ref{eq:3-27})
show that the bias coefficients $c_{ijqlm}$ is the expansion
coefficients of $N_\mathrm{pk}$ by orthogonal polynomials,
\begin{equation}
  \label{eq:b-12}
  N_\mathrm{pk}(\nuc,\bm{x}) = \bar{N}_\mathrm{pk}(\nuc)
  \sum_{i,j,q,l,m}
  \frac{(-1)^q\,q!\,l!\,\Gamma(3/2)\,\Gamma(5/2)\,(2m+1)}
  {i!\,j!\,\Gamma(q+3/2)\,\Gamma(l+5/2)}
  {\sigma_0}^i\,{\sigma_1}^{2q}\,{\sigma_2}^{j+2l+3m}\,
  c_{ijqlm}(\nuc)\,
  H_{ij}(\nu,J_1)\,L^{(1/2)}_q\left(3\eta^2/2\right)\,
  F_{lm}\left(5J_2, J_3\right).
\end{equation}
See Refs.~\cite{GPP12,LMD16} for the original motivation behind the
introduction of these polynomials.

\section{\label{app:coeff}
Proof of Eq.~(\ref{eq:3-30})
}

In this Appendix, we derive the formula of Eq.~(\ref{eq:3-30}) from
the original definition of Eq.~(\ref{eq:3-27}). The calculation is
quite similar to those of Ref.~\cite{BBKS}.

Taking into account the property of Eq.~(\ref{eq:3-28}), the
definition of the bias coefficients, Eq.~(\ref{eq:3-27}), reduces to
\begin{equation}
  \label{eq:c-1}
  c_{ijqlm}(\nuc) = 
  \frac{\chi_q}
  {{\sigma_0}^i\,{\sigma_2}^{j+2l+3m}}
  \frac{
    \int d^{10}y\,\mathcal{P}(\bm{y})\,
    N_\mathrm{pk}(\nuc)\,H_{ij}(\nu,J_1)\,
    F_{lm}\left(5J_2, J_3\right)
  }{
    \int d^{10}y\,\mathcal{P}(\bm{y})\,
    N_\mathrm{pk}(\nuc)
  }.
\end{equation}
Because the integrands in the rhs are rotationally invariant, the
angular degrees of freedom do not contribute to the integrals.
We define invariant variables,
\begin{equation}
  \label{eq:c-2}
  x \equiv \lambda_1 + \lambda_2 + \lambda_3, \quad
  y \equiv \frac{1}{2}(\lambda_1 - \lambda_3), \quad
  z \equiv \frac{1}{2}(\lambda_1 - 2\lambda_2 + \lambda_3),
\end{equation}
where $\lambda_1,\lambda_2,\lambda_3$ are eigenvalues of $-\,\zeta_{ij}$
with a descending order ($\lambda_1\geq\lambda_2\geq\lambda_3$). With
these variables, we have
\begin{equation}
  \label{eq:c-3}
  J_1 = x,\quad
  J_2 = 3y^2 + z^2, \quad
  J_3 = z^3 - 9y^2 z.
\end{equation}
The peak number density of Eq.~(\ref{eq:2-19}) reduces to
\begin{equation}
  \label{eq:c-4}
  N_\mathrm{pk}(\nuc) =
  \frac{2}{\sqrt{3}{R_*}^3}
    \Xi(\nu,\nu_\mathrm{c})\, 
    \delta_\mathrm{D}^3(\bm{\eta})\,
    (x-2z) \left[(x+z)^2-(3y)^2\right]
    \Theta(y-z) \Theta(y+z)\Theta(x-3y+z).
\end{equation}
The probability distribution function of the variables $\nu,x,y,z$ are
given by \cite{BBKS}
\begin{equation}
  \label{eq:c-5}
  \mathcal{P}(\bm{y}) d^{10}y \propto
  \mathcal{N}(\nu,x)
  \left|y\left(y^2 - z^2\right)\right|
  e^{-5(3y^2+z^2)/2}\,
  d\nu\,dx\,dy\,dz,
\end{equation}
besides distributions of $\bm{\eta}$ and angular degrees of freedom.
Multiplying Eqs.~(\ref{eq:c-4}) and (\ref{eq:c-5}), and substituting
the resulting expression into Eq.~(\ref{eq:c-1}), the
Eq.~(\ref{eq:3-30}) is proven.

\section{\label{app:third} Analytic expression of perturbed
  correlation function of peaks }

In this Appendix, the functions $\xi_\mathrm{pk}^{(N)}(r)$ in
Eq.~(\ref{eq:3-41}) for $N=1,2,3$ are explicitly given in terms of the
generalized correlation function, $\xi^{(n)}_l(r)$. The derivation is
straightforward: Eqs.~(\ref{eq:3-25a})--(\ref{eq:3-25c}) are squared
and substituted into Eq.~(\ref{eq:3-23}), and formulas of
Eqs.~(\ref{eq:3-34}) and (\ref{eq:3-37}) are applied. The result of
$N=3$ is derived by a use of the software package,
\textsl{Mathematica}.

For $N=1$, we have
\begin{equation}
  \xi_\mathrm{pk}^{(1)}(r)
  = {b_{10}}^2 \xi^{(0)}_0(r)
  + 2 b_{10}b_{01} \xi^{(2)}_0(r)
  + {b_{01}}^2 \xi^{(4)}_0(r).
  \label{eq:d-1}
\end{equation}
For $N=2$, we have
\begin{align}
  \xi_\mathrm{pk}^{(2)}
  &= {b_{20}}^2 \left(\xi^{(0)}_0\right)^2
    + 4 b_{20}b_{11} \xi^{(0)}_0 \xi^{(2)}_0
    + 2 {b_{11}}^2 \xi^{(0)}_0 \xi^{(4)}_0
    + 2 \left(b_{20}b_{02} + {b_{11}}^2 + \frac{2}{3} {\chi_1}^2\right)
    \left(\xi^{(2)}_0\right)^2
    + 4 b_{11}b_{02} \xi^{(2)}_0 \xi^{(4)}_0
    + 4 b_{20} \chi_1 \left(\xi^{(1)}_1\right)^2
    \nonumber\\
  &\quad
    + 8 b_{11} \chi_1 \xi^{(1)}_1 \xi^{(3)}_1
    + \left( {b_{02}}^2 + \frac{4}{5} {\omega_{10}}^2 \right)
      \left(\xi^{(4)}_0\right)^2
    + 4 \left(b_{02} + \frac{4}{5}\omega_{10}\right)\chi_1
      \left(\xi^{(3)}_1\right)^2
    + 4 \left(b_{20}\omega_{10} + \frac{2}{3} {\chi_1}^2 \right)
      \left(\xi^{(2)}_2\right)^2
    + 8 b_{11} \omega_{10} \xi^{(2)}_2 \xi^{(4)}_2
    \nonumber\\
  &\quad
    + 4 \left(b_{02} + \frac{2}{7} \omega_{10} \right) \omega_{10}
      \left(\xi^{(4)}_2\right)^2
    + \frac{24}{5} \chi_1 \omega_{10} \left(\xi^{(3)}_3\right)^2
    + \frac{72}{35} {\omega_{10}}^2 \left(\xi^{(4)}_4\right)^2.
  \label{eq:d-2}
\end{align}
Finally, for $N=3$, we have
\begin{align}
  \xi_\mathrm{pk}^{(3)}
  &={b_{30}}^2 \left(\xi^{(0)}_0\right)^3
    + 6 {b_{21}} {b_{30}} \xi^{(2)}_0 \left(\xi^{(0)}_0\right)^2
    + 3 {b_{21}}^2 \xi^{(4)}_0 \left(\xi^{(0)}_0\right)^2
    + 12 {b_{10}} {b_{30}} {\chi_1} \left(\xi^{(1)}_1\right)^2 \xi^{(0)}_0
    + 2 \left(3{b_{21}}^2 + 2 {b_{10}}^2 {\chi_1}^2 + 3 {b_{12}}{b_{30}}\right) 
    \left(\xi^{(2)}_0\right)^2 \xi^{(0)}_0
    \nonumber\\
  &\quad
    + 4\left(2 {b_{10}}^2 {\chi_1}^2 + 3 {b_{30}} {c_{10010}}\right)
    \left(\xi^{(2)}_2\right)^2 \xi^{(0)}_0
    + 12 {\chi_1}\left({b_{10}} {b_{12}} + \frac{4}{5} {b_{10}} {c_{10010}}\right)
    \left(\xi^{(3)}_1\right)^2 \xi^{(0)}_0
    + \frac{72}{5} {b_{10}} {\chi_1} {c_{10010}} \left(\xi^{(3)}_3\right)^2 \xi^{(0)}_0
    \nonumber\\
  &\quad
    + 3 \left({b_{12}}^2 + \frac{4}{5} {c_{10010}}^2 \right)
    \left(\xi^{(4)}_0\right)^2 \xi^{(0)}_0
    + 12 \left(\frac{2}{7} {c_{10010}} + {b_{12}}\right) {c_{10010}}
    \left(\xi^{(4)}_2\right)^2 \xi^{(0)}_0
    + \frac{216}{35} {c_{10010}}^2 \left(\xi^{(4)}_4\right)^2 \xi^{(0)}_0
    + 24 {b_{10}} {b_{21}} {\chi_1} \xi^{(1)}_1 \xi^{(3)}_1 \xi^{(0)}_0
    \nonumber\\
  &\quad
    + 12 {b_{12}} {b_{21}} \xi^{(2)}_0 \xi^{(4)}_0 \xi^{(0)}_0
    + 24 {b_{21}} {c_{10010}} \xi^{(2)}_2 \xi^{(4)}_2 \xi^{(0)}_0
    + 2\left(4 {b_{01}} {b_{10}} {\chi_1}^2
    + 3 {b_{12}} {b_{21}} + {b_{03}} {b_{30}}\right) \left(\xi^{(2)}_0\right)^3
    + \frac{4}{3} {b_{30}} {\varpi_{01}} \left(\xi^{(2)}_2\right)^3
    \nonumber\\
  &\quad
    + \left({b_{03}}^2
    + \frac{12}{5} {c_{01010}}^2
    + \frac{14}{225} {\varpi_{01}}^2\right) \left(\xi^{(4)}_0\right)^3
    - 4\left(\frac{12}{7} {c_{01010}}^2
    + \frac{41}{3087} {\varpi_{01}}^2
    - \frac{1}{3} {b_{03}} {\varpi_{01}}
    + \frac{6}{49} {c_{01010}} {\varpi_{01}}\right) \left(\xi^{(4)}_2\right)^3
    \nonumber\\
  &\quad
    + \frac{1296}{8575} {\varpi_{01}}^2 \left(\xi^{(4)}_4\right)^3
    + 4\left(4 {b_{01}} {b_{10}} {\chi_1}^2
    + 3 {b_{30}} {c_{01010}} + 3 {b_{21}} {c_{10010}}\right)
    \xi^{(2)}_0 \left(\xi^{(2)}_2\right)^2
    + 4 {\chi_1}\left(3 {b_{03}} {b_{10}} + 3 {b_{01}} {b_{12}}
    - 2 {b_{01}}^2 {\chi_1} \right) \xi^{(2)}_0 \left(\xi^{(3)}_1\right)^2
    \nonumber\\
  &\quad
    + \frac{48}{5} {\chi_1} \left({b_{10}} {c_{01010}}
    + {b_{01}} {c_{10010}}\right)
    \xi^{(2)}_0 \left(\xi^{(3)}_1\right)^2
    + 8{\chi_1}\left(2 {b_{01}}^2 {\chi_1}
    - \frac{12}{5} {b_{10}} {c_{01010}}
    + \frac{7}{25} {b_{10}} {\varpi_{01}}\right)
    \xi^{(2)}_2 \left(\xi^{(3)}_1\right)^2
    \nonumber\\
  &\quad
    + \frac{72}{5} {\chi_1}\left( {b_{10}} {c_{01010}}
    + {b_{01}} {c_{10010}}\right) \xi^{(2)}_0 \left(\xi^{(3)}_3\right)^2
    + \frac{96}{25} {b_{10}} {\chi_1} {\varpi_{01}} \xi^{(2)}_2 \left(\xi^{(3)}_3\right)^2
    + 6\left({b_{03}} {b_{12}}
    + \frac{4}{5} {c_{01010}} {c_{10010}}\right) \xi^{(2)}_0 \left(\xi^{(4)}_0\right)^2
    \nonumber\\
  &\quad
    + 12 \left({b_{12}} {c_{01010}} + {b_{03}} {c_{10010}}
    + \frac{4}{7} {c_{01010}} {c_{10010}}\right) \xi^{(2)}_0\left(\xi^{(4)}_2\right)^2
    + 4 \left({b_{12}} {\varpi_{01}} - \frac{24}{7} {c_{01010}} {c_{10010}}
    - \frac{6}{49} {c_{10010}} {\varpi_{01}}\right) \xi^{(2)}_2 \left(\xi^{(4)}_2\right)^2
    \nonumber\\
  &\quad
    + 2\left( 6 {b_{03}} {c_{01010}} + \frac{144}{35} {c_{01010}}^2
    - \frac{1}{35} {\varpi_{01}}^2 - \frac{8}{5} {c_{01010}}{\varpi_{01}}
    \right) \xi^{(4)}_0 \left(\xi^{(4)}_2\right)^2
    + \frac{432}{35} {c_{01010}} {c_{10010}} \xi^{(2)}_0 \left(\xi^{(4)}_4\right)^2
    + \frac{144}{49} {c_{10010}} {\varpi_{01}} \xi^{(2)}_2 \left(\xi^{(4)}_4\right)^2
    \nonumber\\
  &\quad
    + \frac{8}{35}
    \left({27} {c_{01010}}^2 + \frac{3}{5}{\varpi_{01}}^2
    \right) \xi^{(4)}_0 \left(\xi^{(4)}_4\right)^2
    +  \frac{48}{49}{\varpi_{01}} \left(\frac{1}{7} {\varpi_{01}}
    + 3 {c_{01010}}\right) \xi^{(4)}_2 \left(\xi^{(4)}_4\right)^2
    \nonumber\\
  &\quad
    + 4 {\chi_1} \left( 3 {b_{10}} {b_{21}}
      + 4 {b_{01}} {b_{30}}- 2 {b_{10}}^2 {\chi_1} \right)
    \left(\xi^{(1)}_1\right)^2 \xi^{(2)}_0
    + 16 {b_{10}}^2 {\chi_1}^2 \left(\xi^{(1)}_1\right)^2 \xi^{(2)}_2
    + 8 \chi_1 \left(3 {b_{10}} {b_{12}} + 3 {b_{01}} {b_{21}}
      - 2 {b_{01}} {b_{10}} {\chi_1}\right)
      \xi^{(1)}_1 \xi^{(2)}_0 \xi^{(3)}_1
    \nonumber\\
  &\quad
    + 32 \chi_1 \left({b_{01}} {b_{10}} {\chi_1}
    - \frac{3}{5} {b_{10}} {c_{10010}}\right) \xi^{(1)}_1 \xi^{(2)}_2
    \xi^{(3)}_1
    + \frac{144}{5} {b_{10}} {\chi_1} {c_{10010}} \xi^{(1)}_1 \xi^{(2)}_2
    \xi^{(3)}_3
    + \frac{144}{5} {b_{10}} {\chi_1} {c_{01010}} \xi^{(2)}_2 \xi^{(3)}_1
    \xi^{(3)}_3
    \nonumber\\
  &\quad
    - \frac{48}{25} {b_{10}} {\chi_1} {\varpi_{01}} \xi^{(2)}_2 \xi^{(3)}_1
    \xi^{(3)}_3
    + 12 {b_{01}} {\chi_1} {b_{21}} \left(\xi^{(1)}_1\right)^2 \xi^{(4)}_0
    + 2 \left( 3 {b_{12}}^2
    + 2 {b_{01}}^2 {\chi_1}^2
    + 3 {b_{03}} {b_{21}} \right) \left(\xi^{(2)}_0\right)^2 \xi^{(4)}_0
    \nonumber\\
  &\quad
    + 4\left(\frac{6}{5} {c_{10010}}^2 + 2 {b_{01}}^2 {\chi_1}^2
    + 3 {b_{21}} {c_{01010}}\right) \left(\xi^{(2)}_2\right)^2 \xi^{(4)}_0
    + 12 b_{01} \chi_1 \left({b_{03}} + \frac{4}{5} {c_{01010}}\right)
      \left(\xi^{(3)}_1\right)^2 \xi^{(4)}_0
    + \frac{72}{5} {b_{01}} {\chi_1} {c_{01010}} \left(\xi^{(3)}_3\right)^2 \xi^{(4)}_0
    \nonumber\\
  &\quad
    + 24 {b_{01}} {b_{12}} {\chi_1} \xi^{(1)}_1 \xi^{(3)}_1 \xi^{(4)}_0
    + 4 \left({b_{21}} {\varpi_{01}} - \frac{12}{7} {c_{10010}}^2
    \right) \left(\xi^{(2)}_2\right)^2 \xi^{(4)}_2
    + \frac{8}{5} b_{01} \chi_1 \left(\frac{7}{5} {\varpi_{01}}
    - 12 {c_{01010}} \right) \left(\xi^{(3)}_1\right)^2 \xi^{(4)}_2
    \nonumber\\
  &\quad
    + \frac{96}{25} {b_{01}} {\chi_1} {\varpi_{01}} \left(\xi^{(3)}_3\right)^2
    \xi^{(4)}_2
    + 24 \left( {b_{21}} {c_{01010}} + {b_{12}} {c_{10010}}\right)
      \xi^{(2)}_0 \xi^{(2)}_2 \xi^{(4)}_2
    - \frac{96}{5} {b_{01}} {\chi_1} {c_{10010}} \xi^{(1)}_1 \xi^{(3)}_1
    \xi^{(4)}_2
    + \frac{144}{5} {b_{01}} {\chi_1} {c_{10010}} \xi^{(1)}_1 \xi^{(3)}_3
    \xi^{(4)}_2
    \nonumber\\
  &\quad
    + \frac{48}{5} b_{01} \chi_1
      \left(3 {c_{01010}} - \frac{1}{5} {\varpi_{01}}\right)
      \xi^{(3)}_1 \xi^{(3)}_3 \xi^{(4)}_2
    + 8 \left( 3 {b_{12}}  {c_{01010}} + \frac{6}{5} {c_{01010}} {c_{10010}}
    - \frac{2}{5} {c_{10010}} {\varpi_{01}}\right) \xi^{(2)}_2 \xi^{(4)}_0
    \xi^{(4)}_2
    + \frac{432}{35} {c_{10010}}^2 \left(\xi^{(2)}_2\right)^2 \xi^{(4)}_4
    \nonumber\\
  &\quad
    + \frac{24}{35}\left( 18 {c_{01010}}^2
    + \frac{17}{49} {\varpi_{01}}^2
    - \frac{24}{7} {c_{01010}} {\varpi_{01}}\right)
      \left(\xi^{(4)}_2\right)^2 \xi^{(4)}_4
    + \frac{288}{35} {c_{10010}}\left( 3 {c_{01010}}
    - \frac{2}{7} {\varpi_{01}}\right)
    \xi^{(2)}_2 \xi^{(4)}_2 \xi^{(4)}_4.
  \label{eq:d-3}
\end{align}

\section{\label{app:localNG} Contributions of local-type
  non-Gaussianity to the peak clustering}

In this Appendix, we outline an example of evaluating the non-Gaussian
contribution of Eq.~(\ref{eq:3-50}) to the peak clustering. The
local-type non-Gaussianity is considered to illustrate the calculation
of this term. In the local-type non-Gaussianity, the bispectrum
$B_\zeta(k_1,k_2,k_3)$ of initial curvature perturbations $\zeta$ is
given by
\begin{equation}
  \label{eq:e-1}
  B_\zeta(k_1,k_2,k_3) =
  \frac{6}{5} f_\mathrm{NL}
  \left[
    P_\zeta(k_1) P_\zeta(k_2) + \mathrm{cyc.}
  \right],
\end{equation}
where $P_\zeta(k)$ is the power spectrum of $\zeta$. In Fourier space,
$\tilde{\delta}(\bm{k})$ and $\tilde{\zeta}(\bm{k})$ are related by
\begin{equation}
  \label{eq:e-2}
  \tilde{\delta}(\bm{k},t) = \mathcal{M}(k,t) \tilde{\zeta}(\bm{k},t),
\end{equation}
where
\begin{equation}
  \label{eq:e-3}
  \mathcal{M}(k,t) =
  \frac{2(1+w)}{5+3w} \frac{k^2 T(k,t)}{a^2 H^2 \Omega} g(t).
\end{equation}
In the above equation, $w=1/3$ in the radiation dominated (RD) era and
$w=0$ in the matter dominated (MD) and the post MD eras. $T(k,t)$ is
the transfer function which is non-unity on sub-horizon scales. The
function $g(t)$ is the suppression factor of the potential in the
post-MD epoch, i.e., $g(t) = D_+(t)/a(t)$ with a normalization of
growth factor $D_+(t) \rightarrow a(t)$ in MD epoch. In the RD, MD,
and after the MD epochs, we have
\begin{align}
  \label{eq:e-4}
  \mathcal{M}(k,t) =
  \begin{cases}
    \displaystyle
    \frac{4}{9} \left(\frac{k}{aH}\right)^2 T_\mathrm{r}(k,t),
    &\mathrm{(RD)},
    \\
    \displaystyle
    \frac{2}{5} \left(\frac{k}{aH}\right)^2 T_\mathrm{m}(k,t),
    &\mathrm{(MD)},
    \\
    \displaystyle
    \frac{2}{5} D_+(t)
    \frac{k^2 T_\mathrm{m}(k,t)}{{H_0}^2\Omega_\mathrm{m0}},
    &\mathrm{(post\ MD)},
  \end{cases}
\end{align}
where $T_\mathrm{r}(k,t) = 3^{3/2}(aH/k)j_1(k/(\sqrt{3}aH))$ is the
transfer function in RD epoch, and $T_\mathrm{m}(k,t)$ is the transfer
function in MD and post-MD epochs. We omit the time-dependence from
the argument of functions and denote $\mathcal{M}(k)$, $T(k)$, etc.~in
the following, although all these functions depends on time, as well
as the power spectrum $P(k)$ and the bispectrum $B(k_1,k_2,k_3)$ etc.

From Eqs.~(\ref{eq:e-1}), (\ref{eq:e-2}), the bispectrum of the
smoothed density field $B_\mathrm{s}(k_1,k_2,k_3)$ is given by
\begin{equation}
  \label{eq:e-5}
  B_\mathrm{s}(k_1,k_2,k_3) =
  \frac{6}{5} f_\mathrm{NL}
  \left[
    \frac{\mathcal{M}_\mathrm{s}(k_3)P_\mathrm{s}(k_1) P_\mathrm{s}(k_2)}
    {\mathcal{M}_\mathrm{s}(k_1)\mathcal{M}_\mathrm{s}(k_2)}
    + \mathrm{cyc.}
  \right],
\end{equation}
where
$\mathcal{M}_\mathrm{s}(k) \equiv \mathcal{M}(k)W(kR)$.
Substituting Eqs.~(\ref{eq:3-25a}), (\ref{eq:3-25b}) and
(\ref{eq:e-5}) into Eq.~(\ref{eq:3-50}), we have an analytic
expression of the $P^\mathrm{NG}_\mathrm{pk}(k)$. The same technique
of subsection \ref{subsec:Evaluation} can also be applied, and the
resulting expression reduces to the form which can be calculated very
fast by using \textsl{FFTLog}. The result is given by
\begin{align}
  \label{eq:e-6}
  P^\mathrm{NG}_\mathrm{pk}(k)
  &=
  c_1(k)
  \int_{\bm{k}_{12}=\bm{k}} c_2(\bm{k}_1,\bm{k}_2)
  B_\mathrm{s}(k,k_1,k_2)
  \nonumber\\
&=
  \frac{6}{5} f_\mathrm{NL}  c_1(k)
  \Biggl\{
  \mathcal{M}_\mathrm{s}(k)
  \int_0^\infty4\pi  r^2 dr\,j_0(kr)
  \left[
  b_{20} \left(A^{(0)}_0\right)^2 + 2b_{11} A^{(0)}_0 A^{(2)}_0
  + b_{02} \left(A^{(2)}_0\right)^2 + 
  2\chi_1 \left(A^{(1)}_1\right)^2 + 2\omega_{10} \left(A^{(2)}_2\right)^2
  \right]
\nonumber\\
  &\hspace{5pc} +
  \frac{2P_\mathrm{s}(k)}{\mathcal{M}_\mathrm{s}(k)}
  \int_0^\infty 4\pi r^2 dr\,j_0(kr)
  \left[
  b_{20} A^{(0)}_0B^{(0)}_0
    + b_{11}\left(A^{(0)}_0 B^{(2)}_0 + A^{(2)}_0
    B^{(0)}_0\right)
\right.
\nonumber\\
  &\hspace{22pc}
\left.
 + b_{02} A^{(2)}_0B^{(2)}_0 + 
  2\chi_1 A^{(1)}_1B^{(1)}_1 + 2\omega_{10} A^{(2)}_2B^{(2)}_2
  \right]
  \Biggr\},
\end{align}
where
\begin{align}
  \label{eq:e-7a}
  A^{(n)}_l(r) &\equiv \int\frac{k^2dk}{2\pi^2}
  k^n j_l(kr) \frac{P_\mathrm{s}(k)}{\mathcal{M}_\mathrm{s}(k)},
  \\
  \label{eq:e-7b}
  B^{(n)}_l(r) &\equiv \int\frac{k^2dk}{2\pi^2}
  k^n j_l(kr) \mathcal{M}_\mathrm{s}(k).
\end{align}


\renewcommand{\apj}{Astrophys.~J. }
\newcommand{\aap}{Astron.~Astrophys. }
\newcommand{\aj}{Astron.~J. }
\newcommand{\apjl}{Astrophys.~J.~Lett. }
\newcommand{\apjs}{Astrophys.~J.~Suppl.~Ser. }
\newcommand{\apss}{Astrophys.~Space Sci. }
\newcommand{\cmp}{Comm.~Math.~Phys. }
\newcommand{\jcap}{J.~Cosmol.~Astropart.~Phys. }
\newcommand{\mnras}{Mon.~Not.~R.~Astron.~Soc. }
\newcommand{\mpla}{Mod.~Phys.~Lett.~A }
\newcommand{\pasj}{Publ.~Astron.~Soc.~Japan }
\newcommand{\physrep}{Phys.~Rep. }
\newcommand{\ptp}{Progr.~Theor.~Phys. }
\newcommand{\ptep}{Prog.~Theor.~Exp.~Phys. }
\newcommand{\jetp}{JETP }


\begin{thebibliography}{10}


  \bibitem{bkp} J. R.~ Bond, L.~Kofman and D.~Pogosyan, \nat, \textbf{380}, 603 (1996).

\bibitem{DJS18} V.~Desjacques, D.~Jeong and F.~Schmidt, \physrep
  \textbf{733}, 1 (2018).
  
   \bibitem{krause16} E.~Krause, T.~Eifler and J.~Blazek, \mnras, \textbf{456}, 207 (2016).
  
\bibitem{Dor70} A.~G.~Doroshkevich, Astrofiz. \textbf{6}, 581 (1970).

\bibitem{Kai84} N.~Kaiser, \apjl \textbf{284}, L9 (1984).

\bibitem{PH85} J.~A.~Peacock A.~F.~Heavens, \mnras \textbf{217}, 805
  (1985).

\bibitem{BBKS} J.~M.~Bardeen, J.~R.~Bond, N.~Kaiser \& A.~S.~Szalay,
  \apj \textbf{304}, 15 (1986).

\bibitem{FWDE88} C.~S.~Frenk, S.~D.~M.~White, M.~Davis and
  G.~Efstathiou, \apj \textbf{327}, 507 (1988).

\bibitem{WMS94} T.~Watanabe, T.~Matsubara and Y.~Suto, \apj
  \textbf{432}, 17 (1994).
  
\bibitem{Lud11}  A.~D.~Ludlow and C.~Porciani, \mnras \textbf{413}, 1961 (2011).

\bibitem{PW84} H.~E.~Politzer and M.~B.~Wise, \apjl \textbf{285}, L1
  (1984).

\bibitem{Ott86} S.~Otto, H.~D.~Politzer and M.~B.~Wise, \prl
  \textbf{56}, 1878 (1986).

\bibitem{Cli87}  J.~M.~Cline, H.~D.~Politzer, S.-J.~Rey and
  M.~B.~Wise, \cmp \textbf{112}, 217 (1987).

\bibitem{Cat88} P.~Catelan, F.~Lucchin, S.~Matarrese, \prl
  \textbf{61}, 267 (1988).  
  
\bibitem{LHP89} S.~L.~Lumsden, A.~F.~Heavens and J.~A.~Peacock, \mnras
  \textbf{238}, 293 (1989); Erratum: \mnras \textbf{245}, 192 (1990).

\bibitem{Col89} P.~Coles, \mnras \textbf{238}, 319 (1989).

\bibitem{RS95} E.~Reg\H{o}s and A.~S.~Szalay, \mnras \textbf{272}, 15
  (1995). 
  
\bibitem{Mat95} T.~Matsubara, \apjs \textbf{101}, 1 (1995).

\bibitem{Des08} V.~Desjacques, \prd, \textbf{78}, 103503 (2008).

\bibitem{DCSS10} V.~Desjacques, M.~Crocce, R.~Scoccimarro and R.~K.~Sheth, \prd, \textbf{82}, 103529 (2010).

\bibitem{Bal15} T.~Baldauf, V.~Desjacques and U.~Seljak, \prd,
    \textbf{92}, 123507 (2015).
  
\bibitem{AP90} L.~Appel and B.~J.~T.~Jones, \mnras \textbf{245}, 522 (1990).

\bibitem{PS12} A.~Paranjape and R.~K.~Sheth, \mnras \textbf{426}, 2789
    (2012).

\bibitem{PSD13} A.~Paranjape, R.~K.~Sheth and V.~Desjacques, \mnras
  \textbf{431}, 1503 (2013).

\bibitem{BCDP14} M.~Biagetti, K.~C.~Chan, V.~Desjacques and
  A.~Paranjape, \mnras \textbf{441}, 1457 (2014). 

\bibitem{BCDP16} T.~Baldauf, S.~Codis, V.~Desjacques and C.~Pichon, \mnras
  \textbf{456}, 3985 (2016).

\bibitem{CPP18} S.~Codis, D.~Pogosyan and C.~Pichon, \mnras
  \textbf{479}, 973 (2018).

\bibitem{Sza88} A.~S.~Szalay, \apj \textbf{333}, 21 (1988).

\bibitem{Col93} P.~Coles, \mnras \textbf{262}, 1065 (1993).

\bibitem{LMD16} T.~Lazeyras, M.~Musso \& V.~Desjacques, \prd \textbf{93},
  063007 (2016).

\bibitem{Diz16} A.~Moradinezhad~Dizgah, K.~C.~Chan, J.~Nore{\~n}a,
  M.B~iagetti and V.~Desjacques,
  \jcap \textbf{1609}, 030 (2016).

\bibitem{Mat11} T.~Matsubara, \prd, {\bf 83}, 083518 (2011).

\bibitem{Mat12} T.~Matsubara, \prd \textbf{86}, 063518 (2012).

\bibitem{Mat14} T.~Matsubara, \prd, \textbf{90}, 043537 (2014).

\bibitem{MD16} T.~Matsubara and V.~Desjacques, \prd, \textbf{93},
  123522 (2016). 

\bibitem{SVM16} M.~Schmittfull, Z.~Vlah, P.~McDonald, \prd \textbf{93},
  103528 (2016)

\bibitem{SV16} M.~Schmittfull, Z.~Vlah, 2016, \prd \textbf{94}, 103530
  (2016)

\bibitem{MFHB16} J.~E.~McEwen, X.~Fang, C.~M.~Hirata and J.~A.~Blazek,
  \jcap, \textbf{9}, 015 (2016)

\bibitem{FBMH17} X.~Fang, J.~A.~Blazek, J.~E.~McEwen and C.~M.~Hirata,
\jcap, \textbf{2}, 030 (2017)

\bibitem{PGP09} D.~Pogosyan, C.~Gay, and C.~Pichon, \prd \textbf{80}, 081301
  (2009); \prd \textbf{81}, 129901(E) (2010). 

\bibitem{GPP12} C.~Gay, C.~Pichon, and D.~Pogosyan, \prd \textbf{85},
  023011 (2012).
  
\bibitem{Kac43} M.~Kac, Bull. Amer. Math. Soc., \textbf{49}, 314
  (1943); Bull. Am. Math. Soc., \textbf{49}, 938 (1943).

\bibitem{Ric44} S.~O.~Rice, Bell System Tech. J., \textbf{24}, 46
  (1945). 

\bibitem{BCS08} F.~Bernardeau, M.~Crocce and R.~Scoccimarro, \prd,
  \textbf{78}, 103521 (2008).

\bibitem{Des13} V.~Desjacques, \prd \textbf{87}, 043505 (2013).
  
\bibitem{Ham00} A.~J.~S.~Hamilton, \mnras \textbf{312}, 257 (2000)
  
\bibitem{class11} J.~ Lesgourgues, arXiv:1104.2932.
  
\bibitem{CLASS} D. Blas, J. Lesgourgues, T. Tram, \jcap \textbf{7} 034
  (2011).

\bibitem{Planck2018} Planck Collaboration, arXiv:1807.06209 (2018).
    
\bibitem{Bal13} T.~Baldauf, U.~Seljak, R.~E.~Smith, N.~Hamaus and
  V.~Desjacques, \prd \textbf{88}, 083507 (2013).


\end{thebibliography}

\twocolumngrid

\end{document}